\documentclass[3p,twocolumn,authoryear]{elsarticle}
\usepackage{wrapfig,epsf}

\usepackage{graphicx}
\usepackage[cmex10]{amsmath}
\usepackage{array}
\usepackage{algorithmic}
\usepackage[utf8]{inputenc}
\usepackage{enumitem}
\usepackage{multirow}

\usepackage[figuresright]{rotating}

\newtheorem{algorithm}{Algorithm}[section]

\begin{document}
\begin{frontmatter}

\title{Unanimously Acceptable Agreements for Negotiation Teams in Unpredictable Domains}

\author[upv]{Victor Sanchez-Anguix} \ead{sanguix@dsic.upv.es}
\author[delft]{Reyhan Aydogan} \ead{r.aydogan@tudelft.nl}
\author[upv]{Vicente Julian} \ead{vinglada@dsic.upv.es}
\author[delft]{Catholijn Jonker} \ead{C.M.Jonker@tudelft.nl}

\address[upv]{Departamento de Sistemas Informáticos y Computación\\ Universitat Politècnica de València\\ Camí de Vera s/n, 46022, Valencia, Spain}
\address[delft]{Interactive Intelligence Group\\ Delft University of Technology\\ Delft, The Netherlands}

\begin{abstract}
A negotiation team is a set of agents with common and possibly also conflicting preferences that forms one of the parties of a negotiation. A negotiation team is involved in two decision making processes simultaneously, a negotiation with the opponents, and an intra-team process to decide on the moves to make in the negotiation. This article focuses on negotiation team  decision making for circumstances that require unanimity of team decisions. Existing agent-based approaches only guarantee unanimity in teams negotiating in domains exclusively composed of \textit{predictable and compatible issues}. This article presents a model for negotiation teams that guarantees unanimous team decisions in domains consisting of \textit{predictable and compatible}, and also\textit{unpredictable} issues. Moreover, the article explores the influence of using opponent, and team member models in the proposing strategies that team members use. Experimental results show that the team benefits if team members employ Bayesian learning to model their teammates' preferences.
\end{abstract}


\end{frontmatter}

\section{Introduction}
\label{sec-Introduction}

In the last decade, there has been an increase in the profit earned by electronic commerce systems. This increase has lead to a strong interest of the academic world in researching problems related to e-commerce \citep{ngai02,grieger03,wareham05}. As of today, most e-commerce systems rely on users manually browsing their catalogs and selecting which goods they desire to buy. This task may end up being time consuming and suboptimal in terms of users' preferences, especially as the number of items and services offered on the Web increases.  Therefore, it is necessary to propose mechanisms that helps costumers take better decisions while saving their time efforts.

Agent-based electronic commerce has been proposed as a solution to such problems \citep{guttman98,sierra01,oliveira01,he03}. In an agent-based e-commerce system, autonomous agents act on behalf of their users with the goal of finding and closing satisfactory deals. Automated negotiation is one of the most common approaches when implementing these systems since they allow different electronic parties to reach agreements by exchanging offers and feedback \citep{lomuscio03,nguyen05,buffett07,lau07,chan08}. The benefits of automated negotiation and agent-based e-commerce are many. Being brief, some of the most important include:
\begin{itemize}
 \item As stated, browsing online catalogs for an optimal deal may be time consuming. The state-of-the-art in automated negotiation can complete complex negotiations for multiple issues in less than a few minutes \citep{klein03,williams11,baarslag12}.  
 \item On the one hand, automated negotiation saves the user from having to browse the entire catalog. Additionally, its personal agent is directed by the preferences of the user in the negotiation, which should result in deals that are adjusted to the personal liking of the individual. Personalization has been reported to increase user satisfaction in many computational systems \citep{ball06,liang07}. On the other hand, a dynamic process like automated negotiation allows sellers to adapt their deals to the users' preferences, their current business needs, and their competitor dynamics \citep{he03}.
 \item Agreements achieved by human negotiators, suffer from the \textit{leaving money on the negotiation table} effect \citep{thompson03}. This means that human negotiators are content with current agreements, which are usually suboptimal, when they could have performed much better. Agents in automated negotiation have been reported to provide agreements close to the optimal solution \citep{lai08}.
 \item Compared to centralized and offline approaches (e.g., preference aggregation, recommendation approaches, etc.), automated negotiation is a dynamic and parallel process. For instance, some centralized approaches like preference aggregation are computationally hard especially if the preference space is combinatorial \citep{chevaleyre07}. On the other hand, recommendation approaches only filter prospective deals, but they do not close specific contracts adapted to business needs. Contrarily, automated negotiation can be adapted to current business needs (e.g., concede to gain customers and close fast deals). Additionally, as stated above, team members are also motivated by their own personal interests. Therefore, it is possible that some team members show opportunistic behavior inside the team. In such cases, preference aggregation may be manipulated by exaggerating preferences. Additionally, each parties' preferences are private, therefore making it difficult for the other parties to exploit and manipulate. This latter factor is important, since nowadays most users in electronic applications care about the information they filtrate in systems \citep{taylor03}.
 \end{itemize}

Most negotiation mechanisms proposed for e-commerce settings have focused on solving bilateral or multiparty negotiations where parties are individual agents \citep{faratin98,zeng98,klein03b,nguyen05,coehoorn04,buffett07,lai08, williams11,sanchez-anguix11b,aydogan12}. However, some real life scenarios involve negotiation parties that are not necessarily formed by single individuals. Instead, each party may be formed by more than a single individual. For instance, imagine that a group of travelers wants to go on a holiday together. As a group, they have to negotiate with several travel agencies to get the best travel package for the group. Despite sharing a common goal, each member in the multiplayer party may also be motivated by its own personal interests \cite{mannix05, halevy08}.  Therefore, the group not only faces a possibly difficult negotiation with the travel agency, but it also needs to deal with the conflict present in the group. This type of multi-individual negotiating party has been studied in the social sciences under the name of \textit{negotiation team} \citep{thompson96,thompson01}.

As far as the authors are concerned, multi-individual parties have been overlooked in automated negotiation research. The use of computational models for negotiation teams opens doors for new types of interesting and novel applications in electronic commerce. The inclusion of agent-based negotiation teams allows for e-commerce systems to deploy dynamic deal mechanisms for groups, making of e-commerce a more social system. Classically, when purchasing for groups in e-commerce systems, one representative takes decisions for the whole group. Either he makes decisions according to his own preferences or the group needs to engage in a human negotiation which is usually a costly process due to different schedules, logistics, lack of communication problems or interpersonal conflict \citep{behfar08}. With the inclusion of agent-based negotiation teams these problems are eluded since autonomous agents take decisions jointly while saving time and efforts for their users.

We believe that agent-based negotiation teams could provide potentially interesting new services :

\begin{itemize}
 \item Electronic markets for groups of travelers: Online travel agencies offer their services by means of online catalogs where users can browse different products like flights, hotels, restaurants, activities, etc. The possibilities for travels are vast, and usually a single travel operator may offer thousands of possible trip packages/services. Exhaustively looking through this online catalog for an optimal deal becomes an unfeasible task for humans. Additionally, more often than not, travel is a social activity for groups (e.g., friends, family, young people, etc.). Users can benefit from agent-based negotiation teams since they can exhaustively look for deals while taking the preferences of the group into account and saving efforts. Service providers can also benefit from these models since they could adapt their business strategies in a dynamic way and add a level of personalization that may help to retain customers. Moreover, offering the possibility for groups to close travel deals based on their preferences is a value-added service, that as far as we know, is not currently offered by the industry. As an example of its application, users may indicate to their personal agents their desire to go on a travel together. Then, the agents prepare to negotiate with different travel agencies in order to provide a complete and satisfactory travel package for the users. The fact that the negotiation is carried out automatically by electronic agents also gives room to looking for several alternatives in parallel. Once several trip packages have been negotiated, the personal agents may communicate the agreements to users, who can validate them in the last instance.
 \item Electronic support for agricultural cooperatives: Agricultural cooperatives are supposed to be democratic institutions where groups of farmers join together to save resources for the distribution of their products. One of the main problems of agricultural cooperatives is the principal-agent problem \citep{ortmann07}. Basically, despite being democratic institutions, agricultural cooperatives are managed by a board of directors who take decisions on behalf of the democratic institution. It has been reported in the literature \citep{ortmann07} that dissatisfaction in cooperatives comes from the fact that the goals of members are not aligned with those of the managers. As a novel application for electronic commerce, agent-based negotiation teams may provide support for the processes that are carried out by cooperatives. For instance, the negotiations between agricultural cooperatives and distributors may be supported by an electronic market where the agricultural cooperative is modeled as an agent-based negotiation team. Each member may be represented by an electronic and personal agent that participates in the negotiation team according to the preferences of its owner. This way, if the model is capable of ensuring unanimity with regards to team decisions, it may be possible to avoid the principal-agent problem. Of course, agricultural cooperatives are large institutions and considerable research has still to be done to provide scalable and fair computational models. However, research as the one presented in this article contributes to the obtention of such models in the long term.
\item Groups of energy producers in the smart grid: The smart grid is addressed to be the next generation network for electricity distribution \citep{smart}. In this network, energy generation may come from geographically distributed small generators (e.g., green energy generators) that have to compete with large energy producers. Decisions at the smart grid have to be taken dynamically since energy production and consumption may vary or face unexpected events \citep{ramchurn12}. Recently, agent-based electronic commerce has been proposed as proper paradigm for this scenario due to its dynamic nature and adaptive response \citep{brazier02,lamparter10,morais12,ramchurn12}.  If small generators want to compete with large generators like power plants, they may need to group together and act together as a single generator. Agent-based negotiation teams can give support for the group decision making of small generators in a dynamic environment like the smart grid. For instance, an agent-based negotiation team for the smart grid may decide on different contract attributes like energy price for different time slots, contract duration and cancellation fees with different energy consumers.
\end{itemize}

The applications described above present benefits for electronic commerce systems. However, there are still several issues that need to be solved for deploying real applications based on agent-based negotiation teams due to the novelty of the topic. One of the main issues that should be addressed when designing agent-based negotiation team models is unanimity. The authors argue that, whenever it is possible, it is desirable for the final agreement with the opponent to be unanimously acceptable for all of the team members. When the members of the negotiation team are going to interact in the long term, the intra-team strategy should avoid one or some of the team members being clearly at disadvantage (e.g., unacceptable deal) with respect to the other team members. In the first place, the aforementioned situation may end up in users perceiving unfairness, which may affect commitment to the decision, group attachment, and trust \citep{korsgaard95}. And second, but not the least important, users that are not satisfied with agreements found automatically may end up leaving the electronic commerce application.

The existing approaches  \citep{sanchez-anguix11,sanchez-anguix12,sanchez-anguix12b} have focused on achieving unanimously acceptable agreements for negotiation domains exclusively comprised by \textit{predictable and compatible} issues among the team members.  An issue is \textit{predictable and compatible} if the preference order over issue values is the same for team members and this fact is known from the domain (e.g., price in a team of buyers). While some e-commerce domains are exclusively composed by these issues, many domains also contain issues whose preferential ordering over issue values is not known from the domain (i.e., \textit{unpredictable} issues). For instance, it is difficult to predict from a set of cities which ranking represents the preferences of a traveler, which can diverge from the preferences of other travelers.

This article advances the state of the art in agent-based electronic commerce in two different ways. Firstly, it introduces a new model for agent-based negotiation teams, which could support dynamic negotiations for groups of autonomous agents representing their users. Secondly, the present model is capable of assuring that the final agreement is unanimously acceptable for all of the team members in domains that contain both \textit{predictable and compatible} and \textit{unpredictable} issues. We propose an intra-team protocol in which a team mediator helps team members to reach unanimously acceptable decisions. Furthermore, we propose two negotiation strategies for team members: a basic negotiation strategy based on concession tactics and a negotiation strategy using Bayesian learning to model  teammates' and opponent's preferences for \textit{unpredictable} issues. The model is capable of outperforming state-of-the-art approaches for agent-based negotiation teams. We describe our general framework in Section \ref{Sec-NegoFramework} and the intra-team protocol that allows team members to reach unanimity in Section \ref{Sec-Intra}. After that, we propose two negotiation strategies for team members in Section \ref{Sec-TeamMembers} and we explain why unanimity is guaranteed among team members in Section \ref{Sec-Unanimity}. After analyzing the experiments in Section \ref{Sec-Experiments}, we relate our work to existing approaches and discuss future lines of work in Section \ref{Sec-Discussion}.

\section{Overview of the Negotiation Framework}
\label{Sec-NegoFramework}

Let $A$ represent a negotiation team consisting of $|A| = M$ different team members and a trusted team mediator $med_{A}$, and let $a \in A$ represent a team member in negotiation team $A$. Let $op$ represent the opponent party of the negotiation team. The negotiation between team and opponent is carried out in a bilateral fashion, using an alternating-offers protocol \citep{rubinstein82}. In this protocol, one of the two parties is the initiating party and sends the first offer to the other party or responding party. The responding party receives the offer and decides whether or not he/she accepts the offer. Accordingly, she or he may accept the current offer or send a counter-offer. If the responding agent sends a counter-offer, the initiating party has to decide whether he/she accepts the counter-offer or not. If the counter-offer is rejected, the process is repeated in a turn-taking fashion until a deal is mutually accepted (successful negotiation) or one of the parties decides to quit the negotiation since its deadline has been reached (failed negotiation). Concerning inter-party communications, the team mediator interacts with the opponent by sending team's proposals and transmitting opponent decisions to team members. The team mediator plays a key role since it coordinates the team members and helps them reach unanimously acceptable deals.

Let $X$ be the object under negotiation, $j \in \{1,...,n\}$ be the issues under negotiation,  $D_{j}$ be the negotiation domain or valid values for issue $j$ and $x_{j} \in D_{j}$ represent a valid value for issue $j$. Each agent's preferences are represented by means of a private additive utility function. We assume that there is no preferential interdependency among negotiation issues; that is, the valuation given to a certain issue does not affect preference on the valuation of other negotiation issues. The utility function for an agent in our framework can be formalized as follows:
\begin{equation}
\label{Eq-utility}
		U(X)=w_{1} V_{1}(x_{1}) + w_{2} V_{2}(x_{2})+...+w_{n} V_{n}(x_{n}).
\end{equation}
where $w_{j}$ represents the importance given to issue $j$ by the agent, and $V_{j} : x_{j} \rightarrow [0,1]$ is a scoring function for issue $j$ that gives the score that the agent assigns to an issue value $x_{j}$. It is assumed that $\overset{n}{\underset{j=1}{\sum}} w_{a,j} = 1$ and $w_{a,j}\geq 0$, and then $U(.)$ is a function scaled in [0,1], where $0$ represents the least desirable negotiation deals, and $1$ represents the most desirable negotiation deals. For agents, $RU \in [0,1]$ represents the reservation utility or the minimum level of utility to consider an agreement as acceptable.

In the proposed framework, private information and bounded rationality are assumed. The former has been introduced above: information regarding agents' preferences is private, and so are the strategies and minimum acceptable values of each agent. This is true even among team members, since prior to the negotiation they do not know any information regarding other teammates' preferences. The only information available is obtained via interactions in the intra-team protocol. The latter refers to the fact that given the limited time, information privacy, and limited computational resources, agents cannot calculate the optimal strategy to be carried out during the negotiation. Instead, they employ heuristic strategies that aim to be as good as possible in terms of the achievable utility.

\subsection{Unanimously acceptable agreements}

Each team member $a \in A$ has a reservation utility $RU_{a} \in [0,1]$ that represents the minimum utility that satisfies the team member's need.  Each outcome whose utility is lower than the reservation utility is unacceptable for the team member. As stated along this article, we consider that unanimity in a negotiation team is of extreme importance. An offer is unanimously acceptable for a team $A$ if it is acceptable for all of the team members inside the negotiation team:
\begin{equation}
\forall a  \in A, U_{a}(X) \geq RU_{a}.
\label{eq:unanimity}
\end{equation}

The proposed intra-team strategy will assure that team members only accept those offers that are unanimously acceptable for all the team members and that offers proposed to the opponent are over each team members' reservation utilities, thus, making it unanimously acceptable.

\subsection{Types of negotiation issues among team members}

Among the different negotiation issues that compose the negotiation domain, we consider that there are issues that are \textit{predictable and compatible} among team members and issues that are \textit{unpredictable} among team members.

\begin{figure*}[t]
\centering
  \includegraphics[width=0.7\linewidth]{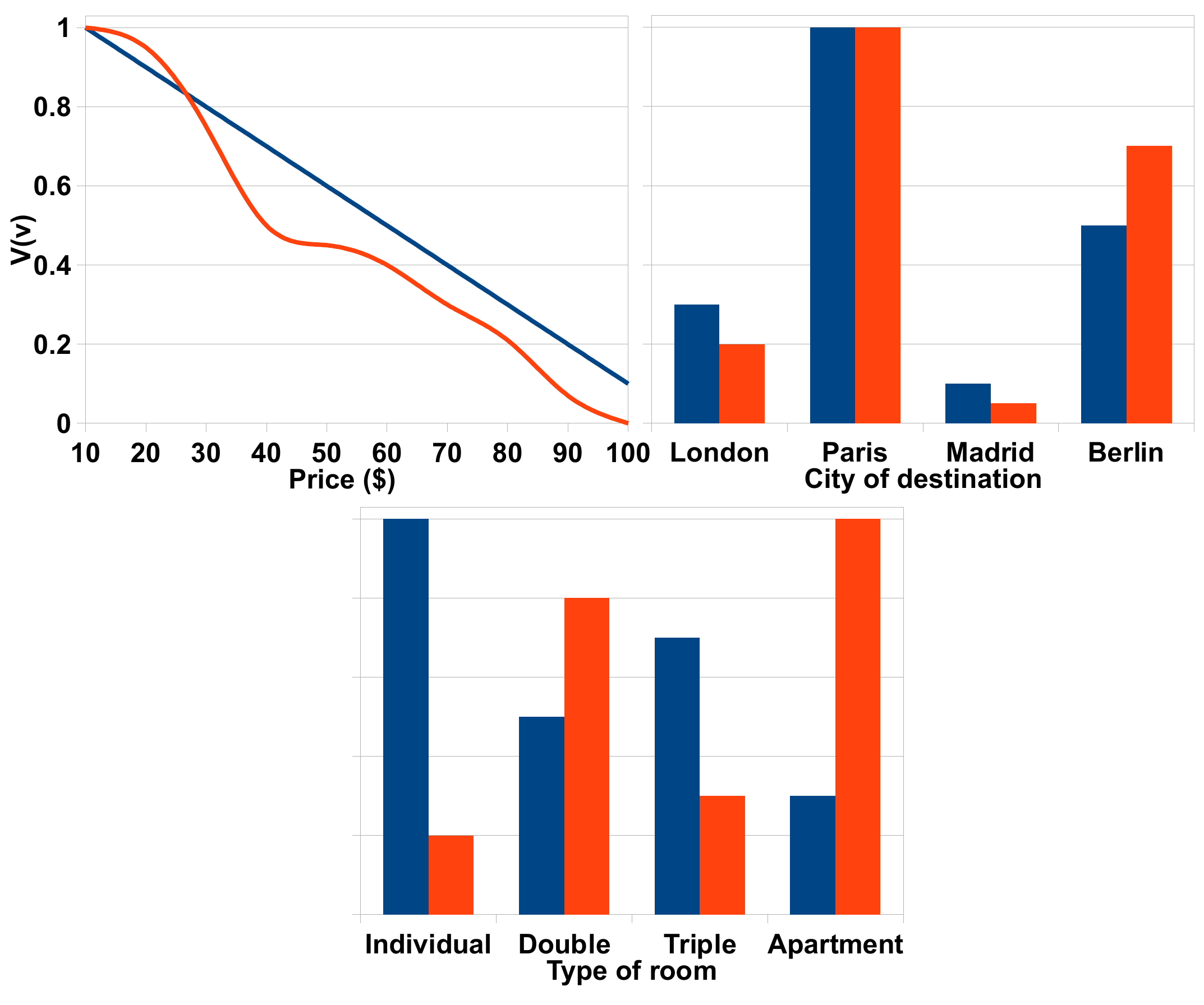}
\caption{Two compatible issues among two agents (top) and a non compatible issue among two agents (bottom).}
\label{fig:comp}
\end{figure*}

  Formally, we can define an issue $j$ with domain $D_{j}$ as compatible among team members if for each possible pair of team members $a, b \in A$ and for each pair of issue values $v1, v2 \in D_{j}$, the following expression is true:
\begin{equation}
 \label{Eq-predictable}
V_{a,j}(v_{2}) > V_{a,j}(v_{1}) \rightarrow V_{b,j}(v_{2}) \geq V_{b,j}(v_{1}).
\end{equation}
Hence, an issue is compatible among team members if one of the team members can increase its utility by selecting a certain issue value with respect to the current assignment, then the rest of team members stay at the same utility or they also increase their utility.  Thus, there is no preferential conflict among issue values between the team members, and there is full potential for cooperation among team members with respect to compatible issues. Figure \ref{fig:comp} shows two examples of compatible issues among two agents (top part) and an example of a non compatible issue (bottom part). As it can be observed, in the case of price (top left), both agents obtain a better valuation when choosing a lower price value with respect to a high price value. Thus, Equation \ref{Eq-predictable} holds and it is a compatible issue for both agents. In the case of the city of destination (top right), the issue is also compatible among the two agents. For any pair of cities, if one of the agents prefers one of the cities with respect to other city, the other agent also holds the same preferential relationship. For instance, both agents prefer Paris to Berlin, Berlin to London, and London to Madrid. However, in the case of the type of room (bottom part), the blue agent prefers an individual room with respect to an apartment, whereas the red agent prefers exactly the opposite. Thus, there is no full potential for cooperation among team members in that negotiation issue since conflict is present.

 The concept of predictability and unpredictability \citep{hindriks10,marsa13} is related to vertical and horizontal issues found in economics literature \citep{stole95}. The definition of predictable issues matches with vertical issues, while the definition of unpredictable issues matches with horizontal issues. From this point on, we will use the concepts of unpredictable/predictable and we will briefly introduce them. An issue is \textit{predictable} for an agent if the preference ordering of issue values is known in the negotiation domain. Therefore, an issue is \textit{compatible and predictable} among team members if the preferences regarding issue values are known in the negotiation domain and increasing the utility of one of the team members by selecting one specific issue value results in other team members staying at the same utility or also increasing their respective utilities. For instance, from the examples in Figure \ref{fig:comp}, one can consider that inside a team of buyers the price is a compatible and \textit{predictable} issue among team members since it is known that all of the buyers prefer low prices to high prices, and reducing the price results in all of the buyers increasing their utility or staying at the same utility. On the other hand, an issue is \textit{unpredictable} among team members if the preference ordering of the issue values cannot be accurately predicted and Equation \ref{Eq-predictable} may not hold for that issue. In the case of Figure \ref{fig:comp}, the city of destination is a compatible issue among the two agents. Nevertheless, in a travel negotiation domain it is not true that all of the travelers will hold the same preference ranking over the issue values and without additional knowledge, the preference ordering may not be predicted accurately. Hence, it is an unpredictable issue. With respect to the type of room, the preference ordering over issue values may vary for the travelers. Moreover, we cannot predict their preference ordering directly, thus making the issue unpredictable.

In this framework, \textit{PR} denotes the set of \textit{predictable and compatible} issues among team members, while \textit{UN} denotes the set of \textit{unpredictable} issues.

\subsection{Forbidden unpredictable partial offers among team members}

We define an unpredictable partial offer $X^{'}$ as a partial offer that has a concrete instantiation of all the unpredictable issues in \textit{UN}. The utility of an unpredictable partial offer is calculated as $U_{a}(X^{'})=\underset{j \in \mbox{\textit{UN}}}{\sum}w_{a,j} V_{a,j}(x_{j})$.

For a team member $a \in A$, an unpredictable partial offer $X^{'}$ will never be part of an acceptable offer (i.e., it will never be an unanimously acceptable offer for the team) when the sum of the utility of $X^{'}$ and the maximum utility that can be obtained from predictable issues $maxPR_{a}=\underset{j \in \mbox{\textit{PR}}}{\sum}w_{a,j} $ is less than its reservation value $RU_{a}$, since any full offer that completes $X^{'}$ is below the reservation utility. For a team member $a$, we refer to the set of unpredictable partial offers that will never be part of an acceptable offer as \textit{forbidden unpredictable partial offers}, $F_{a}$ (see Equation \ref{eq:forbidden1}).

\begin{equation}
F_{a}=\{ X^{'} | U_{a}(X^{'}) + maxPR_{a} < RU_{a}\}
\label{eq:forbidden1}
\end{equation}

It is worth noting that $F_{a}$ does not represent the whole negotiation space that is unacceptable for $a$, but just a portion of it. In fact, some unpredictable partial offers that are not contained in $F_{a}$, can become unacceptable when the agent does not get the value needed from predictable issues. The size of $F_{a}$ may grow as the reservation utility increases. Thus, agents with high reservation utilities are expected to have larger sets of $F_{a}$ than agents with low reservation utilities.


\subsection{Case of Study}
\label{sec-case}
In this article we have employed a case of study (i.e., a negotiation domain) that is extracted from a possible tourism electronic market. The case of study is used to illustrate and test the proposed negotiation framework.

A group of travelers wants to go on a holiday together and arrange their accommodation. The group negotiates with a hotel on the following issues.

\begin{itemize}
 \item \textbf{Price} (\textit{p}): It represents the price per night that each traveler pays to the hotel for the booking service. The value goes from 200\$, which is the minimum rate applicable by the hotel, to 400\$, which is the maximum rate found in the hotel. This negotiation issue is considered to be predictable and compatible among team members since all of the travelers obviously prefer low prices to high prices. Contrarily, the hotel prefers high prices to low prices.
 \item \textbf{Cancellation fee} (\textit{cf}): This issue represents the amount of the final price that each friend pays if the reservation is canceled. Possible values for this negotiation issue go from 0\% to 50\%. This is a predictable and compatible issue among team members since all of the travelers prefer low cancellation fees to high cancellation fees. On the contrary, the opponent prefers high cancellation fees to low cancellation fees.

 \item \textbf{Arranged Foods Included} (\textit{af}): The hotel may also offer some meals included in deal with the travelers. The type of meal plans included are \textit{none}, \textit{breakfast}, \textit{breakfast+lunch}, \textit{breakfast+dinner}, \textit{lunch+dinner}, and \textit{all}. In our negotiation scenario, we have considered that this negotiation issue is unpredictable among team members since preferences of team members on this issue may vary and it cannot be assumed to be same for each member.
 \item \textbf{Type of room} (\textit{tr}): The four travelers can be accommodated in different types of room depending on their preferences. More specifically, the hotel offers \textit{4 individual rooms}, \textit{2 twin rooms}, \textit{1 triple and 1 individual room}, or \textit{1 apartment}. The type of room is an unpredictable negotiation issue among team members.
 \item \textbf{Payment method} (\textit{pm}): The amount of money paid by the travelers may be paid by different methods. The hotel allows for the payment to be made in \textit{cash}, via \textit{credit card}, by \textit{bank transfer}, in a \textit{3 months deferred payment} through the bank, and in a \textit{6 months deferred payment}. This negotiation issue is unpredictable since team members may prefer to choose different payment methods and we cannot predict their preference ordering directly.
 \item \textbf{Room orientation} (\textit{ro}): If possible, the team members can decide upon an orientation for the balcony of their rooms. The different options are \textit{inner garden}, \textit{main street}, \textit{pool}, \textit{sea}, and \textit{outer garden}. This issue is also considered an unpredictable issue among team members.
 \item \textbf{Free amenity} (\textit{fa}): As a token of generosity for booking as a group, the hotel offers one free service to all of the team members. More specifically, the team members can choose between \textit{gym} service, \textit{free wi-fi}, \textit{1 free drink per day}, \textit{1 free spa session}, \textit{pool service}, \textit{cable tv service}, and one \textit{free guided tour}. Since the preferences of team members vary for this issue and no assumption about their preferences can be made, this issue is also considered as unpredictable.
\end{itemize}

To sum up, for this case study we have that \textit{PR}$=\{$\textit{p}$,$\textit{cf}$\}$ and \textit{UN}$=\{$\textit{af}$,$\textit{tr}$,$\textit{pm}$,$\textit{ro}$,$\textit{fa}$\}$ with a total of $4200$ different combinations of discrete issue values (\textit{af},\textit{tr},\textit{pm},\textit{ro},\textit{fa}) and two real issues (\textit{p},\textit{cf}). We assume that the team mediator knows which issues are predictable and can apply an operator that determines the best value for team members from a given set. For unpredictable issues, team members can have different types of valuation functions and the mediator does not know which issue values are better for team members. Each team member may assign different weights (i.e., priorities) to negotiation issues and the opponent's valuation functions and issue weights may be different from those of team members. The team mediator does not know the weights given by agents to the different issues.

\section{Intra-Team Protocol}
\label{Sec-Intra}
\begin{figure*}
 \centering
\includegraphics[width=0.95\linewidth]{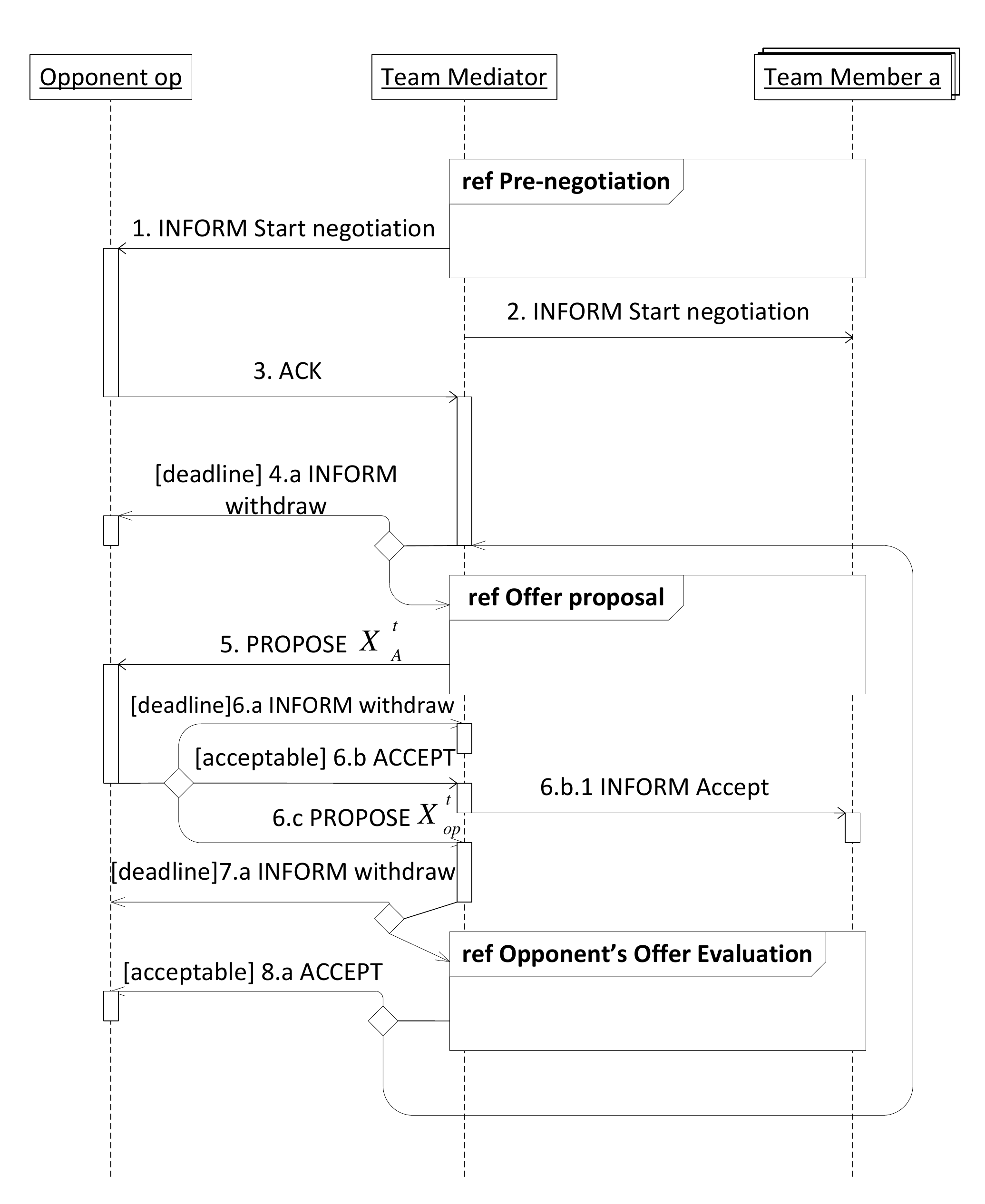}
\caption{Overview of the communications carried out by the team mediator.}
\label{fig:general}
\end{figure*}

In a negotiation involving a negotiation team, the intra-team protocol defines how and when decisions are taken regarding the negotiation. In this framework, we propose an intra-team protocol that is governed by the trusted team mediator $med_{A}$. Basically, the team mediator regulates the interactions that can be carried out among team members and, accordingly, helps team members reaching unanimous acceptable decisions inside the team during the negotiation. The proposed protocol is clearly differentiated into two different phases: \emph{Pre-negotiation} and \emph{Negotiation}. On the one hand, during the pre-negotiation, the mediator helps team members identifying potential offers that are not unanimously acceptable for every teammate. On the other hand, during the negotiation the mediator coordinates the offer proposal mechanism, which is composed of a voting process for unpredictable issues and an iterated building process for predictable issues, and the offer acceptance mechanism for offers that come from the opponent. We describe those phases in a detailed way in the following sections \ref{Sec-PreNego} and \ref{Sec-Nego}. An overview of all of the communications carried out in the negotiation model are depicted in Figure \ref{fig:general}. It specifies the protocols carried out within the team and the communications carried out with the opponent by means of Agent UML \citep{bauer01} sequence diagrams. More detailed views of the intra-team protocols for the pre-negotiation, evaluation opponent's proposals and proposing offers can be observed in Figures \ref{fig:prenegotiationuml}, \ref{fig:negotiation2uml}, and \ref{fig:proposaluml} respectively.

\subsection{Pre-negotiation Phase}
\label{Sec-PreNego}

In the pre-negotiation phase, the mediator coordinates the following intra-team protocol to discover the set of forbidden unpredictable partial offers $F_{A}$ for the team . The set of forbidden unpredictable partial offers for the team, $F_{A}$, is defined as $F_{A}=\{X' | \exists a \in A, X' \in F_{a} \}$. This means that any unpredictable partial offer in $F_{A}$ is never part of an acceptable offer for at least one team member. Thus, these unpredictable partial offers should be avoided for the team since the goal of the negotiation model is reaching unanimously acceptable agreements.

\begin{figure*}
 \centering
 \includegraphics[width=0.75\linewidth]{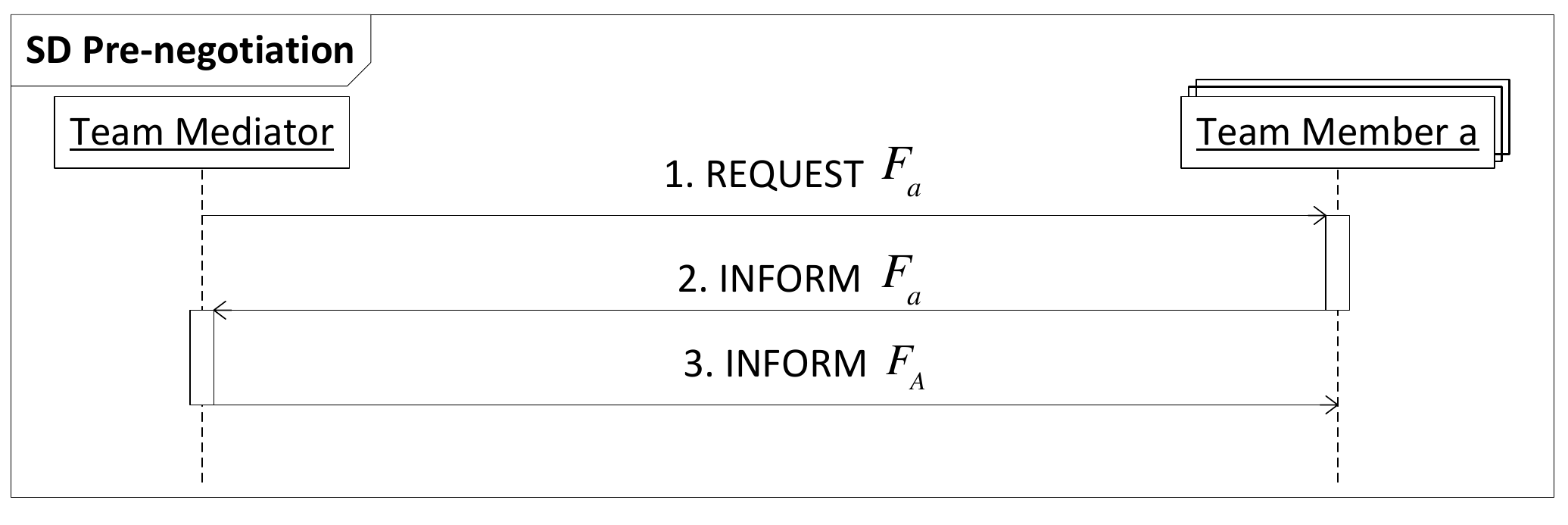}

\caption{Overview of the intra-team protocol carried out during the pre-negotiation}
\label{fig:prenegotiationuml}
\end{figure*}

 A formal description of the pre-negotiation protocol is presented in Figure \ref{fig:prenegotiationuml}. The picture describes the protocol by means of Agent UML sequence diagrams. According to the proposed protocol, the team mediator initiates the pre-negotiation phase by asking each team member $a$ to calculate its own set of forbidden unpredictable partial offers $F_{a}$ (message 1 in Figure \ref{fig:prenegotiationuml}). Each team member builds its own (forbidden) set as requested, and it is communicated to the mediator privately (message 2 in Figure \ref{fig:prenegotiationuml}). When the mediator receives the sets from the team members, it aggregates them in order to construct the set of forbidden unpredictable partial offers for team A, $F_{A}=\underset{a \in A}{\bigcup}F_{a}$. Then, the team mediator makes public the list of forbidden unpredictable partial offers of the team $F_{A}$ (message 3 in Figure \ref{fig:prenegotiationuml}). It should be stated that, since any unpredictable partial offer in this set will prevent one of the team members from reaching its reservation utility, the team is not allowed to generate an offer involving any of these partial offers in $F_{A}$. After the team mediator has shared $F_{A}$ with team members, the negotiation phase starts.

The reader may realize that it is possible that during this phase, most of the \textit{unpredictable} partial offers are pruned. In that case, it means that there is little potential for cooperation among team members. This issue can be observed by the team mediator prior to starting any negotiation process. In that case, the team mediator may suggest the team not to negotiate and save the computational resources used in the negotiation. If the team is not static and can be dynamically formed, it may suggest team members to disband the team and look for other potential partners. However, this team/coalition formation\citep{gaston05,rahwan09} is outside of the scope of this work since we focus on studying the performance of the negotiation model. We consider the use of the information provided with forbidden \textit{unpredictable} partial offers for negotiation team formation as a future line of work. In general, combinations of team members that prune a small portion of the space should be more similar among them, and it should be more easy to achieve cooperation.

\subsection{Negotiation Phase}
\label{Sec-Nego}

In the negotiation, two mechanisms are carried out at each round:  a mechanism for deciding to accept/reject the opponent's offer (Evaluation of Opponent's Offer), and a mechanism for proposing an offer to the opponent (Offer Proposal). For the former, a unanimity voting process is employed, while for the latter an offer building process is governed by the team mediator.

\subsubsection{Evaluation of Opponent's Offer}
\label{Sec-OpponentAcceptance}

\begin{figure*}
 \centering
 \includegraphics[width=0.8\linewidth]{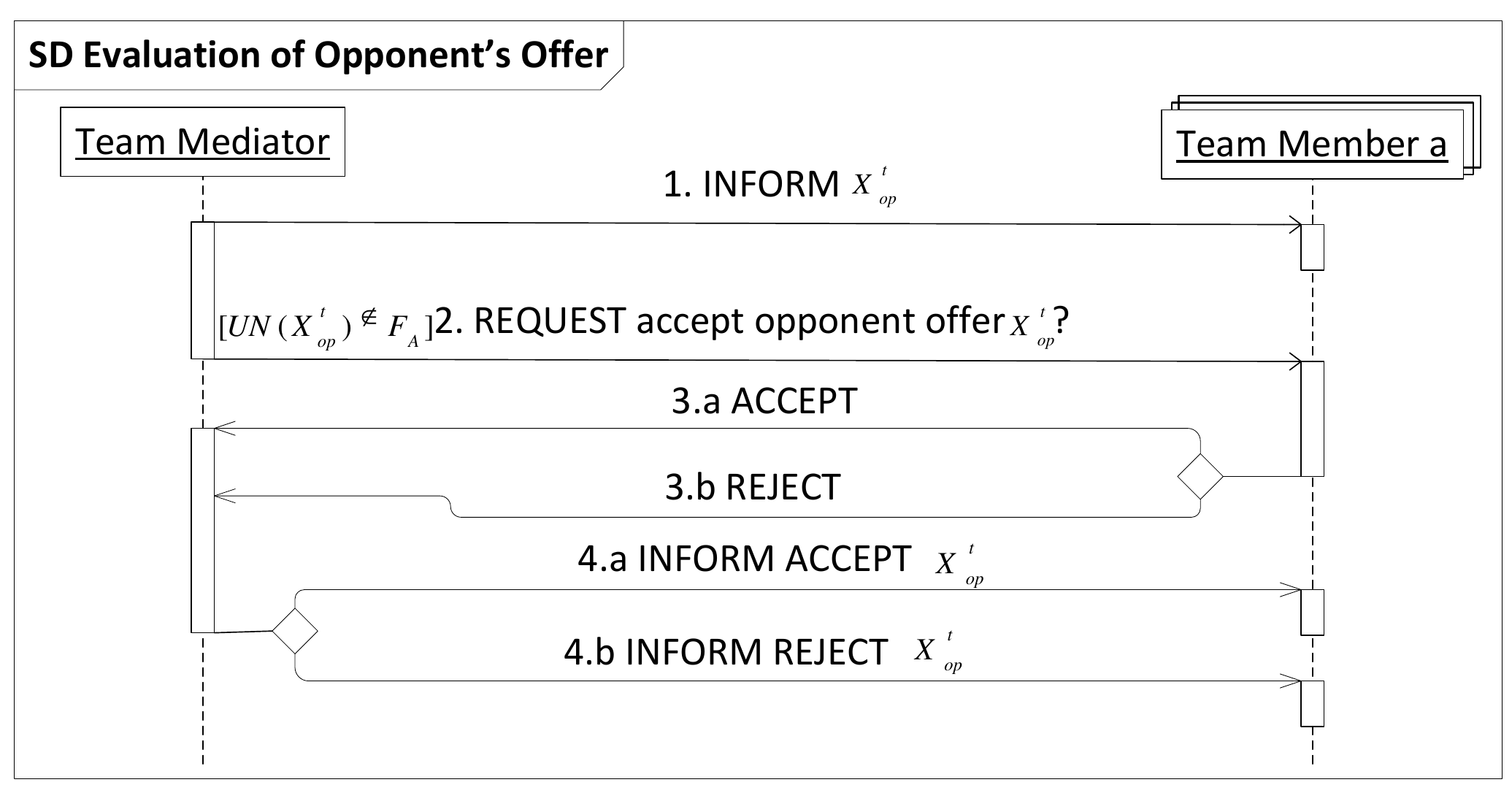}

\caption{Overview of the intra-team protocol employed to evaluate opponent's offers.}
\label{fig:negotiation2uml}
\end{figure*}

This mechanism is carried out each time the team mediator receives an offer from the opponent. Since the main goal of the proposed intra-team strategy is achieving unanimously acceptable agreements for the team, a unanimity voting is carried out to decide whether or not the opponent's offer is acceptable for the team. With this mechanism, as long as one of the team members is not satisfied with the opponent's offer, the offer is not accepted by the team, precluding the team from reaching agreements that are not unanimously acceptable. A formalization of the protocol followed in this mechanism can be observed in Figure \ref{fig:negotiation2uml}. The picture shows the formalization employing sequence diagrams from Agent UML. The intra-team protocol used for this mechanism goes as follows. First, the team mediator receives the offer $X^{t}$ from the opponent at time $t$. If $X^{t}$ involves any forbidden unpredictable partial offer in $F_{A}$, the opponent offer is automatically rejected. However, the opponent's offer is also informed to team members in order to allow each team member to process the new information leaked by the opponent if they see it necessary (message 1 in Figure \ref{fig:negotiation2uml}). Otherwise, if the combination of unpredictable issue values is not in $F_{A}$, in order to see whether the offer is unanimously acceptable for team members, the mediator makes the opponent's offer public among team members and starts an anonymous voting process (message 2 in Figure \ref{fig:negotiation2uml}). Each team member $a \in A$  states to the mediator whether he is willing to accept $X^{t}$ (positive vote) or to reject it (negative vote) at that specific instant (messages 3.a or 3.b in Figure \ref{fig:negotiation2uml}). Since our aim is to guarantee unanimity, the offer is only accepted if all of the team members emit a positive vote (message 4.a in Figure \ref{fig:negotiation2uml}). Otherwise, the offer is rejected and a counter-offer is proposed as explained in Section~\ref{Sec-OfferProposol}.

\subsubsection{Offer Proposal}
\label{Sec-OfferProposol}

\begin{figure*}
 \centering
 \includegraphics[width=\linewidth]{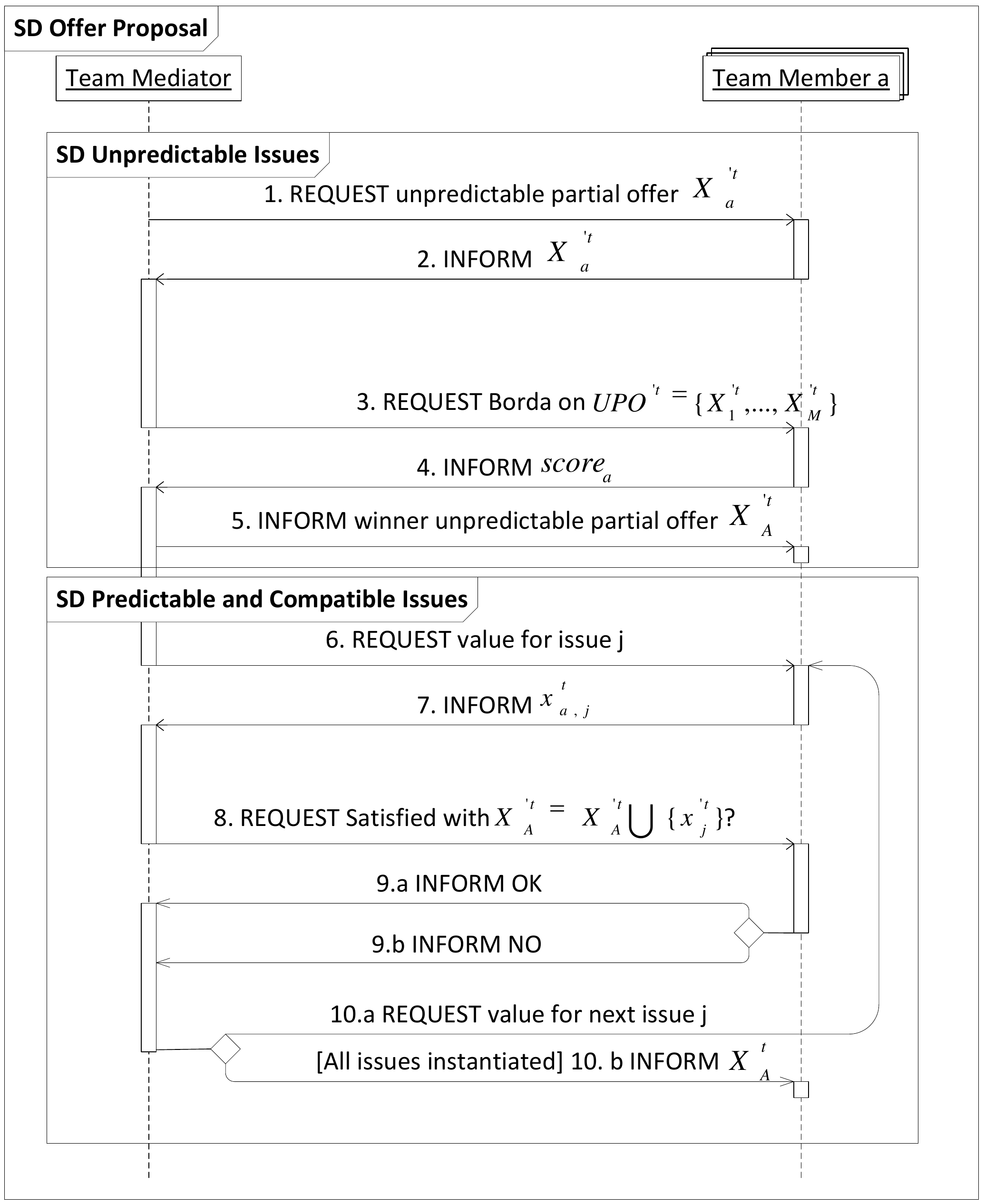}
\caption{Overview of the intra-team protocol carried out to propose offers to the other party.}
\label{fig:proposaluml}
\end{figure*}

Proposing an offer to the opponent is a complex task, since the space of offers may be huge and the preferences of the team members should be reflected in the offer sent to the opponent, and, in our case, the offer sent should be unanimously acceptable for team members. The process is divided into two sub-phases: constructing an unpredictable partial offer, and setting up predictable issues. In both phases, the team mediator acts according to Algorithm \ref{alg:proposeoffer}.  We include another formal description of the interactions between the mediator and a team member during the offer proposal. This information can be found in Figure \ref{fig:proposaluml}, which depicts the intra-team protocol specified in and Agent UML sequence diagram.

\begin{itemize}

 \item \textbf{Constructing an unpredictable partial offer:} The first step is to propose an unpredictable partial offer, a partial offer which has all of the unpredictable issues instantiated. Since team members know from $F_{A}$ the list of unpredictable partial offers that will not result in unanimously acceptable offers under any circumstance, any offer proposed by the team should avoid being constructed from unpredictable partial offers found in $F_{A}$. The method used to propose offers to the opponent relies on the fact that unpredictable issues are those where intra-team conflict may be present, whereas there is full potential for cooperation in predictable and compatible issues. Hence, in order to build an offer to be sent to the opponent, it seems more appropriate to jointly set unpredictable issue values first and then, depending on the remaining needs of team members, allow team members to set compatible and predictable issues as they require for reaching their demands. The proposed mechanism for the first part, proposing an unpredictable partial offer, is based on voting and social choice. The voting process goes as follows.

\begin{enumerate}
  \item The mediator asks each team member to anonymously propose one unpredictable partial offer $X_{a}^{'t}$ (message 1 in Figure \ref{fig:proposaluml}).
  \item Each team member privately sends its proposal to the mediator, who gathers all of the proposals in a list that will be later sent to team members. If any unpredictable partial offer proposed by $a$ is contained in $F_{A}$, the mediator automatically ignores this proposal (message 2 in Figure \ref{fig:proposaluml}).
  \item Once all of the proposals have been gathered, the mediator makes public the list of proposal  $\text{UPO}^{'t}$ among team members and opens a Borda scoring process \citep{nurmi10} on proposed candidates (message 3 in Figure \ref{fig:proposaluml}).
  \item Each team member anonymously scores candidates and sends the scores to the team mediator (message 4 in Figure \ref{fig:proposaluml}).
  \item The team mediator sums up scores and selects the winner candidate with the highest score $X_{A}^{'t}$, making it public among team members (message 5 in Figure \ref{fig:proposaluml}). This candidate, an unpredictable partial offer, will be the base for the full offer that will be sent to the opponent.
\end{enumerate}

\item \textbf{Setting up predictable and compatible issue values:} Once unpredictable issues have been set, it is necessary to set predictable and compatible issues to construct a complete offer. As it has been stated along this article, there is full potential for cooperation among team members in these issues since increasing the utility of one of the team members by selecting one issue value will result in the other team members staying at the same utility or increasing their utility. Obviously, the selected unpredictable partial offer will not satisfy equally the needs of all the team members. Nevertheless, team members can make use of predictable and compatible issues to satisfy their remaining needs while not generating conflict inside the team. To complete the partial offer $X_{A}^{'t}$, an iterative mechanism that we proposed in \citep{sanchez-anguix12} is used to build the final offer issue per issue. The mechanism follows an order for predictable and compatible issues that is constructed by the mediator at each round according to the history of the opponent's concessions.  The rationale used to build this order is that the opponent would concede less on those predictable issues more important for him in the first negotiation rounds, whereas it would concede more on those predictable issues that are less important. Thus, the order established by the team mediator attempts to order predictable and compatible issues in ascending order of importance for the opponent. The general idea behind this ordering is attempting to satisfy team members' demands with those predictable issues less important for the opponent first. The order is updated as new information becomes available from the offers sent by the opponent. Based on this order, the iterative mechanism goes as follows.

\begin{enumerate}[resume]
 \item The mediator selects the first predictable issue $j$ and asks team members, given the current partial offer $X_{A}^{'t}$, the necessary value $x_{j}$ for $j$ to get as close as possible to their current demands (message 6 in Figure \ref{fig:proposaluml}).
 \item Accordingly, each team member $a$ informs the mediator privately about the most convenient value $x_{a,j}$ for that issue (message 7 in Figure \ref{fig:proposaluml}). To decide the final value $x_{j}$ for the issue $j$, the trusted mediator aggregates agents' opinions (since the issue is predictable) by means of a function that, for team members, returns the most preferred issue value from a given set ($best(.)$).  After deciding the value $x_{j}$, $X_{A}^{'t}$ is updated with $x_{j}$.
 \item The mediator asks the team whether or not the new partial offer is already satisfactory at round $t$ (message 8 in Figure \ref{fig:proposaluml}).
 \item Each team member emits an affirmative response if the current partial offer covers its current demands and a negative response if it still has not covered its demands (message 9.a or 9.b in Figure \ref{fig:proposaluml}). Those agents that agree with the current state of $X_{A}^{'t}$ leave the iterative mechanism for this offer since they already are satisfied with the current partial offer. The process steps back to the selection of the next issue.
 \item The process continues until all of the predictable issues have been set or until all of the team members have left the iterative mechanism. In the latter case, the remaining issues are set attempting to maximize the opponent's preferences. Once the offer is complete, it is announced among team members and sent to the opponent (message 10.b in Figure \ref{fig:proposaluml}).
\end{enumerate}
\end{itemize}

\begin{algorithm}

\label{alg:proposeoffer}
Pseudo-algorithm for the offer construction from the point of view of the mediator. Send (\textit{message} $\longrightarrow$ \textit{condition} ) means that \textit{message} is sent to every agent that fulfills \textit{condition}
\begin{algorithmic}[1]
\STATE
\STATE /*Proposing an unpredictable offer*/\;
\STATE Send (REQUEST $X_{a}^{'t} \longrightarrow \forall a \in A)$\;
\STATE Receive (INFORM $X_{a}^{'t}\longleftarrow \forall a \in A$)\;
\STATE $\text{UPO}^{'t}=(\underset{a \in A}{\bigcup}X_{a}^{'t})-F_{A}$\;
\STATE Send (REQUEST Borda on $\text{UPO}^{'t} \longrightarrow \forall a \in A)$\;
\STATE Receive (INFORM $score_{a} \longleftarrow \forall a \in A)$\;
\STATE $X_{A}^{'t}=\underset{X' \in \text{UPO}^{'t}}{\text{argmax}} \underset{a \in A}{\sum} score(a,X')$\;
\STATE Send (INFORM $X_{A}^{'t} \longleftarrow \forall a \in A)$\;
\STATE $order=build\_predictable\_order();$ $\;\;\;A'=A$\;
\STATE
\STATE /*Setting predictable issues*/\;
\FORALL{$j \in order$}
\STATE Send (REQUEST value for $j$ $\longrightarrow \forall a | a\in A'$)\;
\STATE Receive (INFORM $x_{a,j}$ $\longleftarrow \forall a| a\in A'$)\;
\STATE $x_{j}=best( \{x_{a,j} | a\in A' \})$\;
\STATE $X_{A}^{'t}=X_{A}^{'t} \bigcup \{x_{j}\}$\;
\STATE Send (REQUEST Satisfied with $X_{A}^{'t}$? $\longrightarrow \forall a\in A'$)\;
\FORALL{$a\in A'$}
 \STATE Receive (INFORM $ac'_{a}(X_{A}^{'t})$ $\longleftarrow$ $a$)\;
 \IF{$ac'_{a}(X_{A}^{'t})=true$}
   \STATE $A'=A'-\{a\}$\;
 \ENDIF
\ENDFOR
\IF{$A'=\emptyset$}
 \STATE break;
 \ENDIF
\ENDFOR
\FORALL{$j \in order \wedge issue\_not\_set(j)$}

\STATE $x_{j}=maximize\_for\_opponent(j)$\;
 \STATE  $X_{A}^{'t}=X_{A}^{'t} \bigcup \{x_{j}\}$\;
\ENDFOR
\STATE $X_{A}^{t}=X_{A}^{'t}$\;
\end{algorithmic}

\end{algorithm}

\section{Team Members' Strategies}
\label{Sec-TeamMembers}
The team mediator defines the coordination mechanisms inside the team. However, each team member's internal strategy has a great effect on team dynamics.  In this article, we propose two types of strategy for team members. According to the first strategy (i.e., our basic team member), the team member only proposes unpredictable partial offers based on its own utility. In the second strategy, team members model the preferences of the team and the opponent on unpredictable partial offers. Then, in the mechanism employed to set unpredictable partial offers, each team member selects the candidate that guarantees that it can reach its current aspirations at time $t$, and maximizes the probability of being acceptable for the opponent and the team. The learning mechanism employed by these team members is Bayesian learning (i.e., Bayesian team member).

\subsection{Basic Strategy for Team Members}
\label{Sec-BasicTeamMember}

Since negotiations are time-bounded in our framework, we consider that team members have to perform some kind of concession if an agreement is to be found. For this purpose we have designed basic team members as agents whose demands are controlled by an individual and private concession strategy. More specifically, the concession strategy for a team member $a \in A$ is based on time-based tactics $s_{a}(t)$ \citep{faratin98,lai08}. It estimates the utility demanded by $a$ at time $t$ by using the formula in Equation~\ref{Eq-concession}, where $RU_{a}$ is its reservation utility, $T$ is the negotiation deadline, and $\beta_{a}$ is the concession speed, which determines how fast the agent's demands are lowered towards $RU_{a}$.

\begin{equation}
 s_{a}(t)=1-(1-RU_{a})\times (\frac{t}{T})^{\frac{1}{\beta_{a}}}
 \label{Eq-concession}
 \end{equation}

Based on this concession tactic, each team member participates in the intra-team protocol with their demands regulated by his private concession tactic.
Next, we define how team members take their decisions: evaluating the opponent's offer, and proposing an offer for the opponent.

\subsubsection{Evaluation of Opponent's Offer}

Given an offer $X^{t}$ proposed by the opponent at instant $t$, the team member emits a positive vote in the unanimity voting process if it reports a utility greater than or equal to its current demands $s_{a}(t)$. Otherwise, a negative vote is emitted.

\begin{equation}
\label{op_ac}
ac_{a}(X^{t}) = \left\lbrace
	  \begin{array}{l l}
	     true & \mbox{if } s_{a}(t) \leq U_{a}(X^{t}) \\
	     false & \mbox{otherwise}
	  \end{array}
	  \right.
\end{equation}

 \subsubsection{Offer Proposal}

As documented in Section \ref{Sec-Nego}, team members interact at three points during the offer proposal. First, they propose an unpredictable partial offer to the team mediator. Since each team member $a$ has its demands regulated by a time-based tactic, when proposing an unpredictable partial offer to the mediator at instant $t$, the proposed unpredictable partial offer $X_{a}^{'t}$ fulfills:

 \begin{equation}
 X_{a}^{'t} \notin F_{A} \wedge  (U_{a}(X_{a}^{'t}) + maxPR_{a} \geq s_{a}(t))
 \label{eq:partialof}
 \end{equation}

Hence, agent $a$ selects an unpredictable partial offer which is not forbidden inside the team (since it will be ignored by the team mediator) and whose utility allows him to achieve or surpass its current demands at time t. This way, the team member assures that if its proposed unpredictable partial offer is the winner of the Borda voting process, it can reach its current demands. However, one should be aware that many unpredictable partial offers may fulfill Equation \ref{eq:partialof}. Therefore, it is necessary to select one of them as the proposed candidate. Being our basic team member, from the set of partial offers that fulfill Equation \ref{eq:partialof}, a team member selects one of the candidates randomly.

The second time that a team member interacts with the team mediator is for scoring unpredictable partial offers that have been proposed by team members. For scoring candidate partial offers in the Borda voting process, a basic team member orders the candidates according to the partial utility reported by each of the candidates.  That is, the team member assigns the highest score to the partial offer whose utility is the highest for itself, and the second highest score to the partial offer whose utility is the second best one, and so forth.

Finally, team members also interact with the mediator during the mechanism used to set \textit{predictable and compatible} issues.  When team members are asked about a value for issue $j$, each team member communicates anonymously the value $x_{a,j}$. The value is the one that, given the current partial offer $X^{'t}_{A}$, gets the utility of the new partial offer as close as possible to its current aspiration $s_{a}(t)$. Taking normalized additive utility functions, it can be calculated as:
\begin{equation}
\label{eq:bid1}
 x_{a,j}= \left\lbrace
       \begin{array}{l }
        \underset{x\in D_{j}}{\mbox{argmax } V_{a,j}(x)}  \mbox{\;\;\;\;if } U_{a}(X_{A}^{'t})+w_{a,j} \leq s_{a}(t) \\
        \underset{\scriptsize \begin{array}{l}x\in D_{j} \\ \wedge \\ w_{a,j} V_{a,j}(x) \geq (s_{a}(t) - U_{a}(X_{A}^{'t})) \end{array}    } {\mbox{argmin } V_{a,j}(x)}  \mbox{    otherwise }
       \end{array}
	\right.
\end{equation}
where $s_{a}(t)$ is the utility demanded by the agent $a$ at round $t$, $U_{a}(X_{A}^{'t})$ is the utility reported by the current partial offer, and $w_{a,j}V_{a,j}(x)$ is the weighted utility reported by the value demanded by the agent. The value asked for issue $j$ is the closest one to the current demands of the agent. On the one hand, if the agent cannot reach its current demands by just setting issue $j$, it asks for the value that reports the highest utility. On the other hand, if the agent can reach or surpass its current demands by setting $j$, it asks for the value that makes the new partial offer the closest to the current demands. In the same iterative process, team members still have to declare whether or not they are satisfied with the different partial offers that are constructed. Team members follow a similar criterion to the method proposed to determine if an opponent offer is acceptable at $t$. Basically, a partial offer is acceptable for an agent $a$ at $t$ if it reports a utility greater than or equal to the aspiration level marked by its concession strategy:
\begin{equation}
\label{eq:accept}
 ac'_{a}(X_{A}^{'t}) = \left\lbrace
	  \begin{array}{l l}
	     true & \mbox{if } U_{a}(X_{A}^{'t}) \geq s_{a}(t) \\
	     false & \mbox{otherwise}
	  \end{array}
	  \right.
\end{equation}
where $true$ indicates that the partial offer is acceptable at its current state for agent $a$, and $false$ indicates the opposite.

\subsection{Bayesian-based Strategy for Team Members}
\label{Sec-BayesianTeamMember}

The Bayesian-based negotiation strategy for a team member is based on modeling the team's (as a whole) and its opponent's preferences on unpredictable issues, and acting accordingly.  For this purpose,  two Bayesian models are employed to predict if unpredictable partial offers are acceptable for both teammates and the opponent. One of the Bayesian models is employed to capture the preferences of the team on unpredictable issues, whereas the other is used for capturing the preferences of the opponent on unpredictable issues. The strategy used to evaluate the opponent's offer is the same as the one described in the basic strategy.

\subsubsection{Bayesian Learning}

Bayesian learning is a probabilistic learning method based on Bayes' theorem \citep{norvig03}. Given a certain set of hypothesis $H$ and some observation $e$, Bayesian learning attempts to compute the probability $p(h|e)$ that a certain hypothesis $h$ is true after observing $e$. In our case, we want to determine whether or not the proposed offer will be acceptable for the opponent (or the team) (H=$\{acc,\neg acc\}) $ given a certain unpredictable partial offer ($e=X^{'t}$) where $acc$ stands for ``acceptable'' and $\neg acc$ stands for ``unacceptable''.

Since we assume that there is no interdependence among negotiation issues, we can consider that each negotiation issue contributes individually to the acceptability of an offer/unpredictable partial offer. Thus, applying Bayes' theorem under independence assumption we have:
 \begin{equation}
p(acc | X^{'t}) = \frac{p(acc) \underset{j \in \mbox{\textit{UN}}}{\prod} p(x_{j} | acc)}{ \underset{H \in \{acc,\neg acc\}}{\sum}p(H) \underset{j \in \mbox{\textit{UN}}}{\prod} p(x_{j} | H)  }
\end{equation}
where $p(acc)$ is the prior probability for an unpredictable partial offer to be acceptable, $p(\neg acc)$ is the prior probability for an unpredictable partial offer to be non-acceptable, and $p(x_{j} | acc)$ is the conditional probability that assuming $X^{'t}$ as acceptable, it has the value of the $j$ issues instantiated to $x_{j}$.

We consider positive examples $S_{acc}$ as those examples that correspond to the acceptable hypothesis ($acc$) and negative examples $S_{\neg acc}$ as those examples that correspond to the not acceptable hypothesis ($\neg acc$).  For the opponent's model, we employ unpredictable partial offers that have appeared in opponent's offers as positive examples, and unpredictable partial offers that appear in offers rejected by the opponent as negative samples. For the team's model, we use $F_{A}$ and those opponent's offers rejected by team members as the set of negative examples. Winners in the Borda votings (i.e., unpredictable partial offers contained in offers sent to the opponent) are considered as positive examples. For computing $p(x_{j} | h)$, we use the proportion between the number of times that $x_{j}$ appears in hypothesis $h$ ($acc$ or $\neg acc$) and the total number of examples for $h$:
\begin{equation}
 p(x_{j} | h) = \frac{\#\{x_{j} \in S_{h} \}}{|S_{h}|}
\end{equation}

The reasons for employing Bayesian learning are varied. The most important one is that it allows online updating of the model as new samples become available. This is important in a process like negotiation, where at each interaction new information becomes available regarding the opponent's/teammates' preferences. If a computationally expensive learning mechanism was used, it would not be possible to include the new information in the model as it becomes available. Furthermore, the learning mechanism is computationally cheap since it mainly involves counting. This is also important in a real application since it allows for simultaneous negotiation threads to exist, which should be maintained with different opponents to look for the best alternatives in an electronic marketplace.

\subsubsection{Offer Proposal}

Bayesian models are employed to help in the selection of the unpredictable partial offer that is proposed to the other team members. Bayesian team members propose at $t$ unpredictable partial offers in the set defined in Equation \ref{eq:partialof}. Bayesian models help to select a candidate from that set.

However, it is reasonable to think that in the first interactions Bayesian model do not accurately represent other agents' preferences. For that purpose, a team member invests part of the negotiation time $t_{exp}$ in exploring the negotiation space and collecting information regarding the opponent's and the team's preferences. As long as the negotiation process has not surpassed $t_{exp}$, the team member just selects randomly one of candidate unpredictable partial offers as basic team members do. Meanwhile, the Bayesian models are continuously updated with the new information that becomes available during the negotiation. After reaching the time threshold, the team member starts to use Bayesian models in order to select the unpredictable partial offer to be proposed to the mediator during the offer proposal phase. The heuristic used in the selection of the candidate is proposing an unpredictable partial offer that is both acceptable for the team and the opponent. The model has an additional parameter named $p_{esc}$. It represents the probability of avoiding the Bayesian proposal model and using the random proposal model as described in the basic team member model when the negotiation time has gone beyond $t_{exp}$. This parameter is included in the model in order to: (i) further explore the negotiation space; (ii) escape from local optima induced by inaccurate Bayesian models (e.g., wrong samples, limited number of samples, etc.). We can formalize the selection as follows:
\begin{equation}
X_{a}^{'t} = \left\lbrace
	  \begin{array}{l}
	    \underset{X \in B}{\mbox{argmax }} \underset{b\in\{A,op\}}{\sum w_{b} p_{b}(acc | X)}  \mbox{\;\;\;if } \left\lbrace \begin{array}{l}rand \leq p_{esc} \\ \wedge \\  t \geq t_{exp}  \end{array}\right.\\
	      \mbox{random\_partial\_offer}(B) \mbox{\;\;\;otherwise} \\
	  \end{array}
	  \right.
\label{eq-bayesoffer}
\end{equation}
where $B$ is the set of candidate unpredictable partial offers that fulfill Equation \ref{eq:partialof}, $rand$ is a random number, $p_{A}(acc | X)$ is the probability for a candidate unpredictable partial offer to be acceptable for the team, $p_{opp}(acc|X)$ is the probability for the candidate unpredictable partial offer to be acceptable for the opponent, and $w_{A}$ and $w_{op}$\footnote{$w_{A}+w_{op}=1$.} represent the weights given to the acceptability of the unpredictable partial offer for the team and the opponent, respectively (i.e., we will refer to them as Bayesian weights). Varying these Bayesian weights allow team members to show different behaviors depending on their inclination to satisfy either the team or the opponent with the unpredictable partial offer.

\section{Provably Unanimously Acceptable Decisions}
\label{Sec-Unanimity}

As stated in the introduction, one of our research goals is proposing a negotiation team model that is able to guarantee unanimously acceptable team decisions. Next, we show that under the assumption of rationality\footnote{Rational agents seek to improve their current welfare. Thus, they would not take actions that lead to utilities below their reservation utilities.}, team members are able to achieve unanimously acceptable final agreements, if an agreement is found. For that matter, let us employ \textit{reductio ad absurdum} (reduction to absurdity).

If $X$ is the final agreement, let us suppose that Equation \ref{eq:unanimity} (unanimously acceptable) is violated in a negotiation: unanimity is not reached because $a$  obtained a utility below its reservation utility.
 \begin{equation}
 \exists a \in A, U_{a}(X^{'t}) + \underset{j \in \mbox{\textit{PR}}}{\sum} w_{a,j}V_{a,j}(x_{j}) < RU_{a}
 \label{eq:notuna}
 \end{equation}

The final agreement is found when (1) team members accept an opponent's offer or (2) the opponent accepts a team's offer. Next, we show that in both cases, Equation \ref{eq:notuna} is never true.
\begin{enumerate}
 \item When the team members accept an opponent's offer, a unanimity voting process has been carried to decide whether or not to accept the final offer. The offer is only accepted if all of the team members have emitted a positive vote. Since a rational agent $a$ would never have incentive to emit a positive vote of the offer reported a utility below its reservation utility, this scenario is never true due to the intra-team mechanism.

 \item When the opponent accepts a team's offer $X$, this offer has been necessarily proposed by the intra-team mechanism mentioned in Section \ref{Sec-OfferProposol}. The offer can be decomposed into an unpredictable partial offer $X^{'t}$ and the instantiation of predictable issues. The team member $a$ is not able to get over its reservation utility if and only if $X^{'t} \in F_{a}$ or when $X^{'t} \notin F_{a}$ and $a$ could not get what it demanded in predictable issues. A rational agent has no incentive to exclude a forbidden unpredictable partial offer $X^{'t}$ when declaring $F_{a}$. Since $F_{A}=\underset{a \in A}{\bigcup} F_{a}$ and the mediator ignores unpredictable partial offers in $F_{A}$, an unpredictable partial offer $X^{'t}$ that forms a team offer is never in $F_{a}$. If $X^{'t} \notin F_{a}$ then the agent can accomplish to satisfy the following expression $U_{a}(X^{'t}) + maxPR_{a} \geq RU_{a}$. Agent $a$ could not get over its reservation value because he could not demand the most of predictable issues. However, when the team mediator aggregates predictable issues inside the team, the team mediator selects the highest value for team members in the list of values proposed by them. This makes possible for $a$ to obtain the maximum utility from predictable issues. Hence, Equation \ref{eq:notuna}  never holds if the negotiation ends with agreement.
\end{enumerate}

Since both possible scenarios are never true under our initial assumption, we have shown by \textit{reduction ad absurdum} that, if a final agreement is found, it is unanimously acceptable among team members. Another issue is the presence of exaggerating agents (i.e., agents that exaggerate to get the most from the negotiation). In our setting, even if team members exaggerate and decide to include in $F_{a}$ unpredictable partial offers that are acceptable but report low utility, or demand more than they need from predictable issues, if a final agreement is found it will be unanimous among team members. However, by doing so, they may be pruning negotiation space and lowering the probability of finding agreements. This is an interesting situation that we plan to study in the future.

\section{Experiments}
\label{Sec-Experiments}

In this section, we explore the behavior of the proposed negotiation model in different scenarios. The proposed framework has been implemented in \textsc{genius} \citep{lin12}, a simulation framework for automated negotiation that allows researchers to test their frameworks and strategies against state-of-the-art agents designed by other researchers. Recently, \textsc{genius} has become a widespread tool that increases its repository of negotiating agents with the annual negotiation competition \citep{baarslag12}.

In order to assess the performance of the proposed negotiation approach, we have performed different experiments. All of the experiments have been carried out in the negotiation domain (or case study) introduced in Section \ref{sec-case}. The first experiment (Section \ref{Sec-exp1}) studies the performance of the proposed model when facing single opponent agent. The comparison is carried out in scenarios with different degrees of team's preference dissimilarity. In the second experiment, we study the performance of our negotiation team model when facing another negotiation team in bilateral negotiations. In the third experiment (Section \ref{exp-2}) we study how the Bayesian weights $w_{A}$ and $w_{op}$, which control the importance given to the preferences of the team and the opponent in the unpredictable partial offer proposed to teammates, impact the performance of the proposed model when team members employ the Bayesian strategy. Finally, we conduct an experiment to study the effect of team members' reservation utility on the performance of the proposed negotiation model (Section \ref{exp-3}).

\subsection{Performance Against a Single Opponent}
\label{Sec-exp1}

In this first set of experiments we study the performance of the proposed negotiation team model when facing a single opponent. The study is carried out with an emphasis on observing if the performance of the team is higher by employing Bayesian team members rather than basic team members. Due to the fact that we are interested in open environments, we study how the team performance varies with team configurations ranging from scenarios where no team members plays the Bayesian strategy (i.e., all Basic team members) to situations where all the team members play the Bayesian strategy. The performance of the team is measured using the average team joint utility\footnote{We consider the joint utility of the team to be the product of the utilities of the team members. Since the utility of an agent is between 0 and 1, the team joint utility tends to be lower as more team members are present.}. As an additional measure of optimality, we also measure the distance to the closest Pareto optimal point. In this case, the Pareto frontier is computed taking into account the team joint utility and the utility of the opponent. Our initial hypotheses were:

\begin{itemize}
 \item \textbf{H1} As more Bayesian team members form the team, the team is able to obtain average team joint utilities that are higher than or equal to those configurations with less Bayesian team members.
 \item \textbf{H2} As more Bayesian team members form the team, the team is able to obtain average opponent utilities that are higher than or equal to those configurations with less Bayesian team members.
\end{itemize}

Since Bayesian team members take the preferences of the team and the opponent into account when deciding which offer is sent to the other part, they should be able guarantee equal or higher average team joint utility and opponent utility than basic team members. But in no case, they should not obtain lower team joint utility and opponent utility. As a consequence of both hypothesis, the distance to the closest Pareto optimal point should be also equal or lower than that obtained by configurations solely composed by basic team members.

In order to compare the proposed model with other models in the literature, we also included the Similarity Borda Voting model (i.e., SBV) \citep{sanchez-anguix11} in our experiment. SBV is a mediated intra-team strategy that is able to guarantee semi-unanimity regarding team decisions. The mediator imposes a unanimity voting process to decide on whether or not to accept the opponent's offer, whilst team members propose offers to be sent to the opponent by means of a similarity heuristic that takes into account the last offer proposed by the opponent, and the last offer proposed by the team. A Borda voting process is used in order to decide on which offer is sent to the opponent. The reason to include this intra-team strategy in our study is due to the fact that it has been documented to achieve similar results to intra-team strategies that guarantee unanimity under certain circumstances for domains solely composed by predictable issues \citep{sanchez-anguix11}. In order to adapt this approach for domains with unpredictable issues, we use a similarity heuristic that uses Euclidean distance for real/integer issues and string matching for other types of issues. Due to the fact that our proposed model guarantees unanimously acceptable agreements and SBV does not, we formulated the following hypothesis:

\begin{itemize}
 \item \textbf{H3} Teams exclusively formed by basic team members and teams exclusively formed by Bayesian team members obtain equal or higher average team joint utility than teams following the Similarity Borda Voting model.
\end{itemize}

The performance of this first experiment is analyzed in three scenarios with different degrees of preference dissimilarity among team members: very similar, average similarity, and very dissimilar preferences. For this reason, we introduce a measure for measuring team members' preference similarity in different scenarios. The authors proposed a method for calculating preference dissimilarity in teams based on the utility difference of offers among teammates \citep{sanchez-anguix11}. The dissimilarity between two teammates $a,b \in A$ can be measured as:
 \begin{equation}
 D(U_{a}(.),U_{b}(.))= \frac{\underset{\forall X}{\sum} | U_{a}(X)-U_{b}(X) |}{\# \mbox{ possible offers}}
 \end{equation}

Due to the fact that a team may be composed of more than two members, it is necessary to provide a team dissimilarity measure. The team dissimilarity measure is calculated as the average of the dissimilarity between all of the possible pairs of teammates. For this experiment, we decided to explore teams whose preferences are dissimilar, teams whose preferences are similar, and teams with an average degree of similarity/dissimilarity (i.e., average similarity). For the scenario of dissimilar preferences, 9 negotiation cases were randomly generated (i.e., a combination of 3 different negotiation teams consisting of four team members with 3 different opponents), while 9 negotiation cases were randomly generated for the similar preferences scenario (i.e., a combination of 3 different negotiation teams consisting of four team members with 3 different opponents) and 12 negotiation cases were randomly generated for the average similarity scenario (i.e., a combination of 4 different negotiation teams consisting of four team members with 3 different opponents). Since we consider that in practice it is less likely to meet extreme cases such as dissimilar or similar teams, we decided to increase the number of negotiation cases in the average similarity scenario.

As for the single opponents, we decided to test the negotiation team models against different families of opponents. More specifically, we followed the categorization of negotiation strategies proposed by Baarslag \emph{et al.} \citep{baarslag11}, which divides negotiation strategies into four categories:
\begin{itemize}
 \item \underline{\textbf{Competitors:}} They hardly concede, independently of opponent behavior.  \textbf{Agent K} is a competitor agent \citep{kawaguchi11,baarslag11} from the 2010 ANAC competition\citep{baarslag12} that adjusts its aspirations (i.e., target utility) in the negotiation process considering to an estimation of the maximum utility that will be offered by the other party. More specifically, the agent gradually reduces its target utility based on the average utility offered by the opponent and its standard deviation.
 \item \underline{\textbf{Matchers:}} They concede when they perceive that the opponent concedes, and they do not concede if they perceive that the other party does concede. \textbf{Nice Tit-for-Tat} is a matcher agent \citep{baarslag11,baarslag13} from the 2011 ANAC competition that reciprocates the other party's moves by means of a Bayesian model of the other party's preferences. According to the Bayesian model, the Nice Tit-for-Tat agent attempts to calculate the Nash point and it reciprocates moves by calculating the distance of the last opponent offer to the aforementioned point. When the negotiation time is reaching its deadline, the Nice TFT agent will wait for an offer that is not expected to improve in the remaining time and accept it in order to secure an agreement.
 \item \underline{\textbf{Conceders:}} They yield independently of the opponent behavior. \textbf{Conceder} is an implementation  of the time-based concession tactics proposed by Faratin \textit{et al.} \citep{faratin98} categorized by Baarslag \textit{et al.} as conceder. For the Conceder agent $\beta_{op}=2$, which leads to large concessions towards the reservation utility in the first rounds.
  \item \underline{\textbf{Inverter:}} They respond by implementing the opposite behavior shown by the other party. \textbf{Boulware} is an implementation  of the time-based concession tactics proposed by Faratin \textit{et al.} \citep{faratin98} categorized by Baarslag \textit{et al.} as inverter. In the case of the Boulware agent, the concession speed is set to $\beta_{op}=0.2$. Hence, the agent concedes very insignificantly during most of the negotiation and it concedes very quickly as the deadline approaches.

\end{itemize}

\begin{table*}
\small
 \centering
 \begin{tabular}{ | c | c |  c | c | c | c | c | c | c | c | c | c | c | }
\cline{1-13}
  \multicolumn{13}{|c|}{Similar} \\ \cline{1-13}
  &  \multicolumn{3}{c|}{Agent K} & \multicolumn{3}{c|}{Nice Tit-for-Tat} & \multicolumn{3}{c|}{Boulware} & \multicolumn{3}{c|}{Conceder} \\ \cline{1-13}
  &  T.  & Op. & D. & T.   & Op. & D. & T.   & Op. & D. & T.   & Op. & D \\
SBV  &  0.181  &  \textbf{0.743} &   0.070 &  0.150        & 0.694 &  0.130  &  0.184          &  \textbf{0.755}   &  \textbf{0.064}      &  0.552         &  0.482  &   \textbf{0.037}       \\
Basic&  \textbf{0.259}    & 0.683 &  0.065 & \textbf{0.173}    &  \textbf{0.760} &  \textbf{0.067}&  \textbf{0.223}     &  0.696 &  0.078    & 0.561     &  0.468 &   \textbf{0.045}   \\
Bayesian& \textbf{0.263} & 0.690 & \textbf{0.058} & \textbf{0.164}   &  0.746 &  0.080  &   \textbf{0.224}     &  0.695 &  0.080  &  0.557     &    0.472 &   \textbf{0.043}    \\
 \cline{1-13}
  \multicolumn{13}{|c|}{Average Similarity} \\ \cline{1-13}
  &  \multicolumn{3}{c|}{Agent K} & \multicolumn{3}{c|}{Nice Tit-for-Tat} & \multicolumn{3}{c|}{Boulware} & \multicolumn{3}{c|}{Conceder} \\ \cline{1-13}
  &  T. J  & Op. & D. & T. J  & Op. & D. & T. J  & Op. & D. & T. J  & Op. & D. \\
SBV  &    0.168       &  \textbf{0.629}  &   0.065  &   0.137       &    0.562      &   0.116     &    0.170        &   \textbf{0.598}   &  0.070   &  0.324         &   0.428    &  0.074    \\
Basic&   0.211     & 0.574 &  0.070  & 0.141    &   \textbf{0.691}   &   \textbf{0.050} &  0.210         &   0.585  &  0.060   &  \textbf{0.386}       &   0.414  &  \textbf{0.052}    \\
Bayesian& \textbf{0.248} & 0.583 &  \textbf{0.034} &  \textbf{0.158}   & 0.669  &  \textbf{0.047}   &   \textbf{0.224}       &   0.574  &  \textbf{0.045}  &  \textbf{0.390}       &   0.414   &  \textbf{0.050}    \\
  \cline{1-13}
  \multicolumn{13}{|c|}{Dissimilar} \\ \cline{1-13}
  &  \multicolumn{3}{c|}{Agent K} & \multicolumn{3}{c|}{Nice Tit-for-Tat} & \multicolumn{3}{c|}{Boulware} & \multicolumn{3}{c|}{Conceder} \\ \cline{1-13}
  &  T. J  & Op. &  D. & T. J  & Op.    &   D.  & T. J  & Op.     &  D.   & T. J  & Op.   &  D.   \\
SBV  &    0.07      &   \textbf{0.522} &   0.168  &  0.160        &     0.457 &      0.157      &   0.128         &   \textbf{0.547}    &  0.110    &   0.257        &    \textbf{0.430}   &  0.110     \\
Basic&   0.174      &   0.397 &  0.180 &   0.184   &   \textbf{0.572}   &   \textbf{0.055}  &   0.254        &   0.505   &   \textbf{0.053}  &   \textbf{0.472}        &    0.367   &  \textbf{0.046}   \\
Bayesian& \textbf{0.209}& 0.457   &  \textbf{0.118}  &   \textbf{0.196}  &   \textbf{0.559}   &  \textbf{0.60}    &  \textbf{0.271}         &  0.489    &    \textbf{0.058}  &  \textbf{0.475}         &  0.367  &    \textbf{0.044}     \\
\cline{1-13}
 \end{tabular}
 \caption{Average team joint utility (T. ), the average opponent utility (Op.), and the average Euclidean distance to the closest Pareto optimal point (D.) for the first set of experiments.}
 \label{res-1}
\end{table*}

The reservation utility of each team member was set to $RU_{a}=0.5$ to represent scenarios where team members have average aspirations. Additionally, for each team member (i.e., basic, Bayesian and SBV) the concession speed was randomly selected from a uniform distribution $\beta_{a}=U[0.5,1]$. In the case of Bayesian members, the time of exploration was set to $t_{exp}=70\%$ and the probability of escape after the exploration phase was set to $p_{esc}=30\%$ \footnote{These values were found to be the best ones after carrying out a grid search over values of $t_{exp}$ and $p_{esc}$ in a subset of test negotiation scenarios}. Therefore, Bayesian models are not used unless a 70\% of the negotiation time (126 seconds) has passed. Initially, we set Bayesian members to care equally about the acceptability of unpredictable partial for the team and the opponent $w_{A}=w_{op}=0.5$.

Following the type of setting used in the annual agent competition, the negotiation time was set to $T=180$ seconds. Each opponent strategy was faced against each negotiation team model in every possible negotiation case. A total of 20 repetitions were done per negotiation case in order to capture stochastic variations in negotiation strategies. Therefore, $ 3 \times 3\times 3 \times 4 \times 20 = 2160$ (team preference profiles $\times$  opponent preference profiles $\times$  team negotiation models $\times$  opponent strategies $\times$  repetitions)   negotiations were simulated in the similar scenario, 2160 negotiations were simulated in the dissimilar scenario, and $2880$ negotiations were simulated in the average similarity scenario.

\begin{figure*}
 \centering
 \includegraphics[width=1\linewidth]{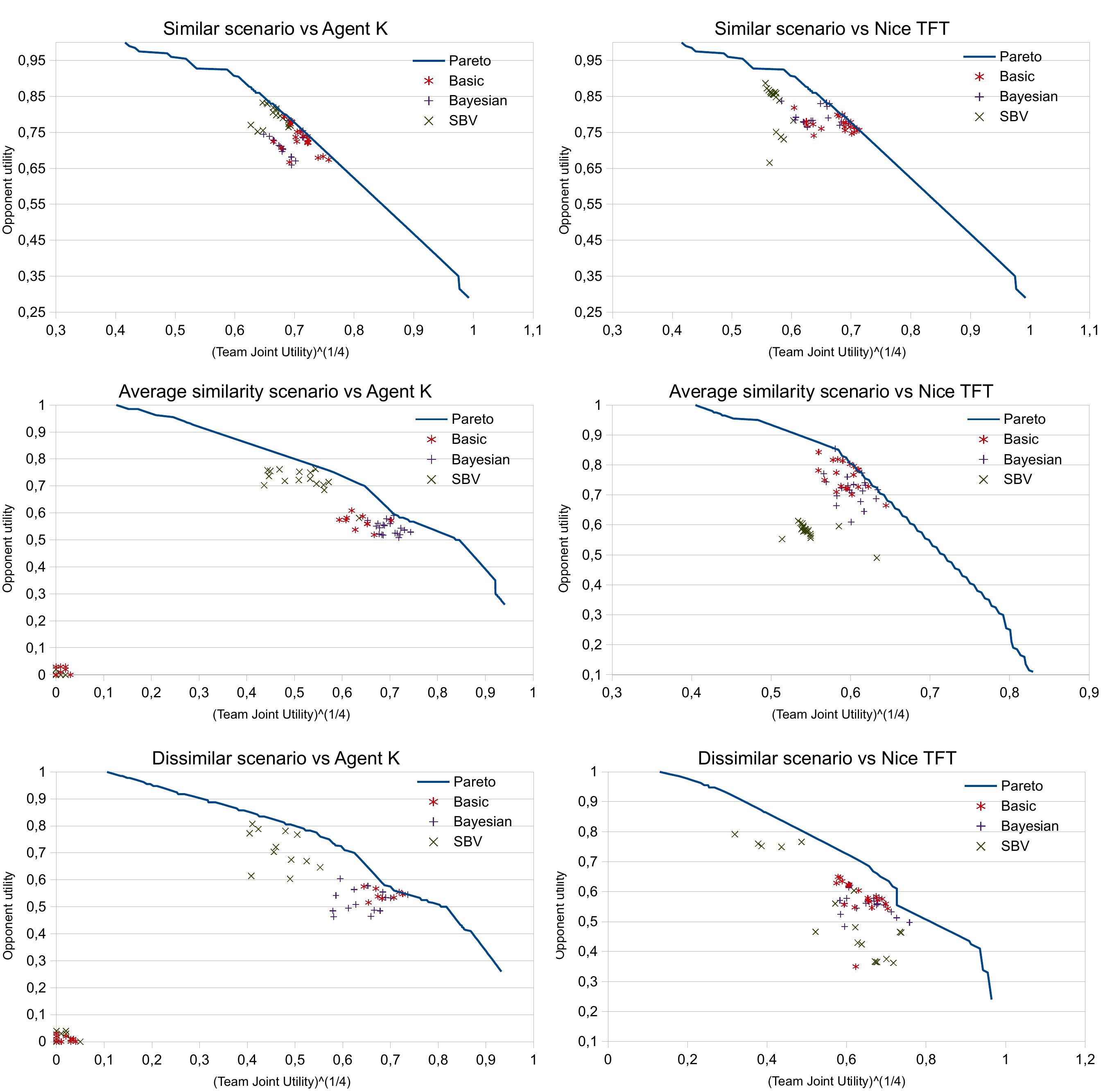}
\caption{Examples of agreements and their distance to the Pareto frontier}
\label{fig:pareto}
\end{figure*}

Table \ref{res-1} shows the average team joint utility and opponent utility for the cases where all of the team members either play the Basic strategy, the Bayesian model, or the team employs the SBV team negotiation model. It also shows the Euclidean distance to the closest point in the Pareto frontier.
An ANOVA test ($\alpha=0.05$) with a Bonferroni post-hoc analysis was carried out to assess statistical differences among the different measures gathered. Those measures that are statistically the best configurations for each column are highlighted in bold style. All of the claims in this experimental section are supported by the ANOVA test with the Bonferroni post-hoc analysis. The average negotiation time taken by each method is included in Table \ref{res-time}. Figure \ref{fig:mixed}  shows the evolution of the average team joint utility and the average opponent utility as more team members play the Bayesian strategy. We have also included some examples of agreements obtained in the different negotiation scenarios and how they relate to the Pareto frontier\footnote{The quadratic root of the team joint utility is taken to convert the results to the same scale (remember that the team joint utility is the product of for team members' utilities)}. These results can be observed in Figure \ref{fig:pareto}. Next, we analyze the results.

\subsubsection{Results for the first hypothesis}

\begin{itemize}
\item \textbf{H1} As more Bayesian team members form the team, the team is able to obtain average team joint utilities that are higher than or equal to those configurations with less Bayesian team members.
\end{itemize}

First, we focus on the situations when all of the team members either play the basic strategy or the Bayesian strategy (Tables \ref{res-1} and \ref{res-time}). It can be observed that when team members' preferences are similar, both types of teams perform equally in terms of the average team joint utility. This result is consistent with H1, since both prove to be statistically equivalent with the ANOVA test with Bonferroni post-hoc analysis. The reason why Bayesian models do not give an advantage over the basic model in the similar scenario can be explained since team members are similar and there is no necessity to carry out team modeling. The distance to the closest Pareto optimal point is also very similar for both team configurations, which can be also explained due to the fact that team modeling is not necessary due to team members' similar preferences.

\begin{figure*}[t]
 \centering
 \includegraphics[width=1\linewidth]{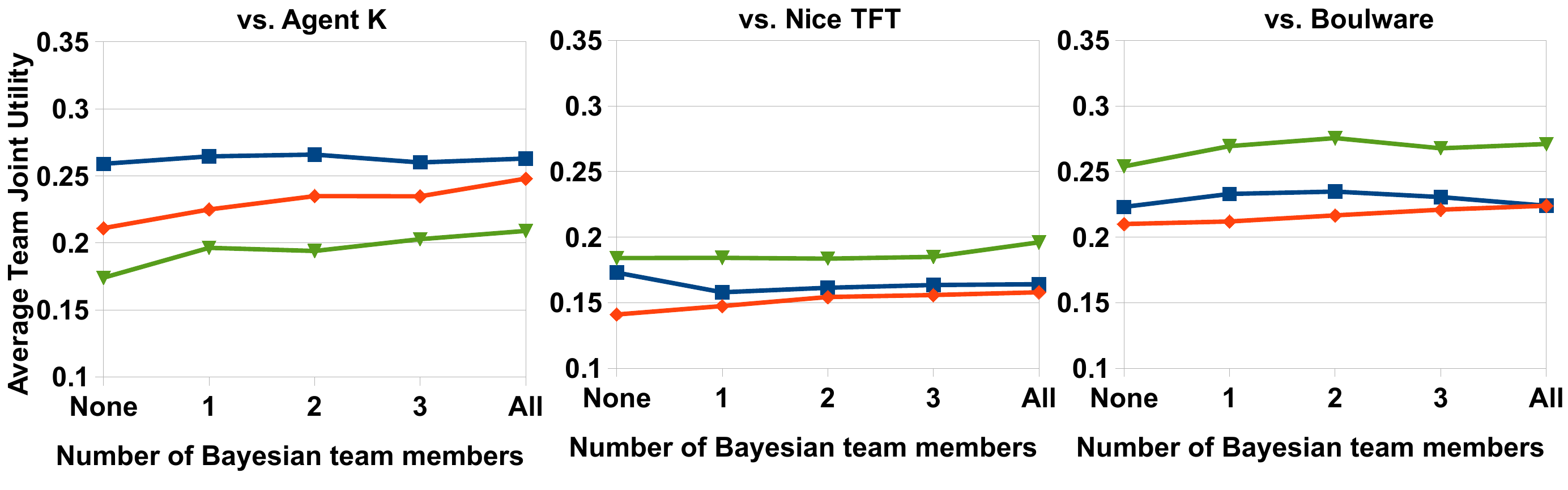}
 \includegraphics[width=1\linewidth]{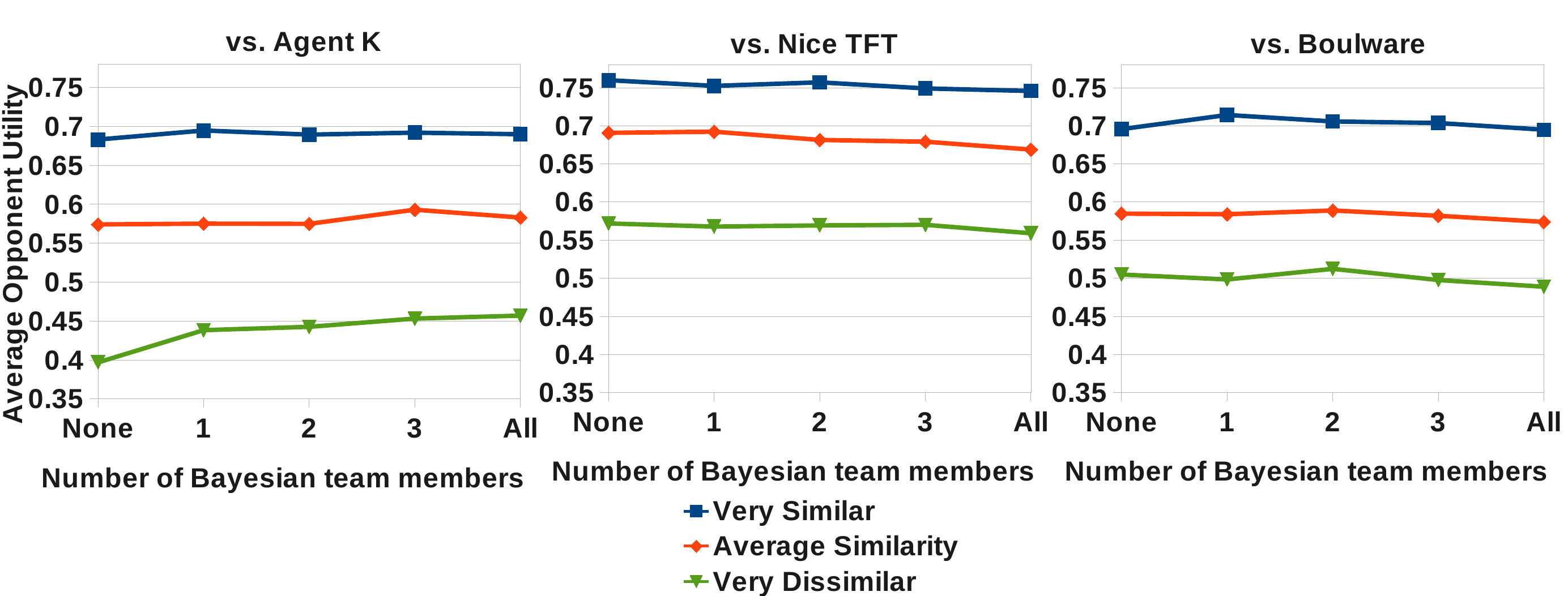}

\caption{Evolution of the average team joint utility and opponent utility as more Bayesian team members are introduced in the team.}
\label{fig:mixed}
\end{figure*}

As conflict is introduced inside the team by making team members' preferences more dissimilar (i.e., \textit{dissimilar and average similarity scenarios}, middle and bottom part of Table \ref{res-1}), it can be observed that usually the team formed exclusively by Bayesian team members gets the statistically highest average team joint utility, which is also coherent with our hypothesis H1 and refines our hypothesis for these scenarios. In this case, the teammates' preferences are no longer similar and some sort of modeling mechanism is needed in order to guide the intra-team negotiation towards agreements that are good for all of the team members. The only exception is found in the conceder case, where the performance in terms of the team joint utility was found to be statistically equivalent among the team exclusively formed by basic team members and the team exclusively formed by Bayesian members. Taking a closer look at the negotiation traces, we observed that, in all of the negotiations, the exploration time $t_{exp}$ was never surpassed. Since the Conceder agent concedes rapidly in the negotiation, the team's demands are also met early. Therefore, Bayesian models do not get to be used. In fact, the average negotiation time against Conceder agents was $61.7$, $76.3$, and $88$ seconds respectively  in the similar scenario (see Table \ref{res-time}),  the average similarity scenario (see Table \ref{res-time}), and the dissimilar scenario (see Table \ref{res-time}). All of them are below the threshold of $126$ seconds delimited by $t_{exp}$. As a result, the team members have not used their Bayesian model while generating their proposals and they are equivalent to the team formed by basic members. This is also consistent with H1, since in no case the team formed by Bayesian members gets statistically lower results than the team formed by basic members.

\begin{table}
\small
 \centering
 \begin{tabular}{|l|l|l|l|l|}
\cline{1-5}
  \multicolumn{5}{|c|}{Similar} \\ \cline{1-5}
  &  K & N. TFT & B. & C. \\ \cline{1-5}
SBV  &  148.7  &  164.3   &  139.4       & 61.0      \\
Basic&  141.5    & 162.5  & 145.4    &  62.8       \\
Bayesian& 142.3 & 165.2 & 144.0   &  61.7           \\
 \cline{1-5}
  \multicolumn{5}{|c|}{Average Similarity} \\ \cline{1-5}
   &  K & N. TFT & B. & C. \\ \cline{1-5}
SBV  &    155.1       &  177.6      &   153.1       &    74.1           \\
Basic&   153.1     & 174.8  & 154.9    &  77.4       \\
Bayesian& 150.1 & 175.0  &  154.7   & 76.3                \\
  \cline{1-5}
  \multicolumn{5}{|c|}{Dissimilar} \\ \cline{1-5}
   &  K & N. TFT & B. & C. \\ \cline{1-5}
SBV  &    163.8     &   175.7  &  156.6        &     73.4     \\
Basic&  162.2     & 176.6   & 160.1   &       87.7         \\
Bayesian& 163.7      &   177.2  &   160.5   &     88.0           \\
\cline{1-5}
 \end{tabular}
 \caption{Average time (seconds) for negotiations in the first set of experiments. K (Agent K), N. TFT (Nice Tit-for-Tat), B. (Boulware), C. (Conceder)}
 \label{res-time}
\end{table}

If we observe the evolution of the average team joint utility in Figure \ref{fig:mixed} \footnote{Results for the Conceder agent are omitted since the Bayesian models are not employed.},  there is a tendency to increase the average team joint utility as more Bayesian members are included (triangle shaped data series on Figure \ref{fig:mixed}) in situations where team members' preferences have an average similarity or they are very dissimilar. This tendency is more pronounced against Agent K (left plot in Figure \ref{fig:mixed}) and Boulware agents (right plot in Figure \ref{fig:mixed}). However, when team members' preferences are very similar the team performance remains at statistically equivalent values (ANOVA test with Bonferroni analysis) for the average team joint utility. The results of these graphics are coherent with our findings in Table \ref{res-1} and H1.

In conclusion, we have found that as more Bayesian team members form the team, the team is able to obtain average team joint utilities that are higher than or equal to those configurations with less Bayesian team members. Being more specific, we have been able to detect that, as long as there is preferential conflict among team members (i.e., average similarity among team members, and very dissimilar team members), and the opponent does not concede early in the negotiation, more Bayesian team members result in higher team joint utility.

\subsubsection{Results for the second hypothesis}

\begin{itemize}
 \item \textbf{H2} As more Bayesian team members form the team, the team is able to obtain average opponent utilities that are higher than or equal to those configurations with less Bayesian team members.
\end{itemize}

For the average opponent utility, the Bayesian team obtained significantly better results than the basic team only in the scenario where the team faces Agent K and team members are dissimilar with regards to their preferences (see Table \ref{res-1}, bottom part). The same pattern is found in Figure \ref{fig:mixed}, where we can observe that the average opponent utility increases as more Bayesian team members are present in the team (first plot in the second row of Figure \ref{fig:mixed}). In other cases, the Bayesian and the basic team obtain statistically equivalent results to each other, and, in some situations, the basic team obtained significantly better results. More specifically, when the team faces Nice Tit-for-Tat, the basic team obtains significantly better results than the Bayesian team. We can also observe this pattern in Figure \ref{fig:mixed}. As more Bayesian members are introduced, the average opponent utility slightly decreases  (middle plot in the second row of Figure \ref{fig:mixed}).

These findings only support partially our hypothesis H2 since we found a set of scenarios where the basic team provides better utility to the opponent (i.e., when facing Nice Tit-for-Tat).  We analyzed the trace of different negotiations against Nice Tit-for-Tat and Boulware opponents. In the former case, we could observe that close to the end of the negotiation the Nice Tit-for-Tat opponent had only sent on average $5$ different unpredictable partial offers in a domain that has $4200$ different unpredictable partial offers. This behavior results in scarce information for any learning mechanism. In the case of negotiations against Boulware agents, one should consider that the Boulware strategies concede only towards the end of the negotiation and, most of the time, the aspirations are high. Thus, most of the samples gathered by the Bayesian classifier when facing Boulware agents correspond to offers with high demands where usually only the best issue values appear. Other issue values do not appear in the samples or they have their frequency misinterpreted with respect to the utility that they actually report. Therefore, the learning mechanism misinterprets the preferences of the opponent and the team formed by Bayesian members is not able to obtain statistically better utility for the opponent than the team formed by basic members.

The  behavior of Nice Tit-for-Tat and Boulware opponents also has a direct consequence on the distance to the closest Pareto optimal point obtained by both team configurations (Bayesian and basic). Despite the fact that the Bayesian configuration is capable of obtaining statistically better results for the team joint utility by learning the preferences of the team, the utility reported to the opponent is usually lower than the one reported by the basic configuration. The only exception to this case are scenarios against Agent K, where the Bayesian configuration obtains a statistically higher utility for the opponent. The inability to model the opponents' preferences in the case of the Boulware and Nice Tit-for-Tat opponents, results in a higher team joint utility (due to team modeling) at the cost of reducing the utility received by the opponent. Hence, there is not an improvement in the distance to the closest Pareto optimal point. These findings can also be observed in some of the examples included in Figure \ref{fig:pareto}.The agreements found by the basic team configuration and the Bayesian configuration tend to be found at the same distance to the Pareto frontier (the Bayesian configuration tending to populate regions with higher team joint utility). The exception to this rule are negotiations against Agent K, where the basic configuration tends to populate regions of no agreement (close to the axis origin).

\subsubsection{Results for the third hypothesis}

\begin{itemize}
 \item \textbf{H3} Teams exclusively formed by basic team members and teams exclusively formed by Bayesian team members obtain equal or higher average team joint utility than teams following the Similarity Borda Voting model.
\end{itemize}

It can be observed that when \textit{team members' preferences are similar} (top part of Table \ref{res-1}), both basic and Bayesian models are statistically equivalent to each other and  better than SBV with respect to the average team joint utility. As conflict is introduced inside the team by making team members' preferences more dissimilar (i.e., \textit{dissimilar and average similarity scenarios}, middle and bottom part of Table \ref{res-1}), the team get statistically lower average team joint utility by employing the SBV model. Basic and Bayesian models outperform SBV with respect to the average team joint utility since they are able to guarantee unanimously acceptable agreements, while SBV does not guarantee such condition. This result supports and refines our hypothesis H3, since, in general, the team formed exclusively by Bayesian team members and the team formed by basic team members obtains statistically higher team joint utilities than the SBV model. There is only one exception to this refinement. The basic team, the Bayesian team and SBV perform statistically equal in terms of the average team joint utility only when the opponent is a conceder and team members' preferences are similar. Since the opponent concedes rapidly in the first rounds, the three models obtain equivalent team joint utility due to the opponent concessions and the fact that conflict is almost nonexistent among team members. However, this finding is still consistent with our initial guess H3.

Regarding the optimality of the solutions found by the model proposed in this article with respect to SBV, it is possible to observe that both Bayesian and basic configurations obtains statistically lower distance to the Pareto frontier than SBV as long as preferential conflict between team members is present (i.e., average similarity and dissimilar scenarios). Only when the team is very similar, SBV gets statistically equal distances to the model proposed in this article. However, despite the fact that the distance to the Pareto frontier may be statistically equal, we can observe in Figure \ref{fig:pareto} that agreements found by SBV tend to populate areas that are closer to the lowest team joint utility. This result can be explained due to the fact that unanimity cannot be guaranteed in the team, and some team members end up with low utility agreements.

In conclusion, we found that  teams exclusively formed by basic team members and teams exclusively formed by Bayesian team members obtain equal or higher average team joint utility than teams following the Similarity Borda Voting model. More specifically, we found that, in general, the results for the team formed by Bayesian team members and the team formed by basic team members obtains statistically higher results than the SBV model. The distance to the Pareto frontier also shows higher quality and more optimal agreements for teams employing the model proposed in this article. This result is important, since it shows that the present model, not only guarantees unanimously acceptable agreements, but it also ensures that better results are obtained with respect to other state-of-the-art team negotiation models.

\subsection{Performance Against Another Team}
\label{Sec-expteam}

In this experiment we analyze the performance of the proposed model when two teams face each other. More specifically, we simulate negotiations where one negotiation team represents a group of four travelers, and the other negotiation team represents the board of managers for a hotel, which consists also of four managers.  The negotiation has a common deadline of $T=180$ seconds. Both parties negotiate with each other by means of the alternating bilateral protocol, but they employ the negotiation team model proposed in this article to coordinate and take team decisions. The goal of the experiment is determining if Bayesian team member improve the performance of the team when it faces another team.

In this setting, we use 4 different team preference profiles to represent the group of travelers and 2 different team preference profiles to represent the board of managers. Since we are interested in open environments, we consider different team configurations from the perspective of strategy profiles: No team member plays the Bayesian strategy (0-0),half of the team members play the Bayesian strategy  in one team  (2-0),  all of the team members play the Bayesian strategy in one team (4-0),  half of the team members are Bayesian in both teams (2-2), all of the team members are Bayesian  in one team and half of them are Bayesian in the other team (4-2), and all of the team members are Bayesian (4-4). For each strategy profile, each group of travelers'  is faced against each board of managers  20 times to capture stochastic variations. Therefore, a total of $4\times 2 \times 20 = 160$ negotiations is carried out per team strategy profile, giving a total of 960 negotiations for this experiment.

Our initial guess is that more Bayesian team members will help to obtain higher team joint utilities due to the learning and proposal mechanism used to take into account the preferences of the team and the preferences of the opponent. More specifically, we formulated the following hypotheses:

\begin{itemize}
 \item \textbf{H4} As long as only one team includes Bayesian team members (configurations 2-0 and 4-0), the average team joint utility for both teams will be higher than the average team joint utility obtained by negotiations where no Bayesian team member participates (configuration 0-0).
 \item \textbf{H5} Those configurations where both teams include Bayesian team members (configurations 2-2, 4-2, and 4-4) will obtain higher average team joint utilities for both teams than configurations where only one team includes Bayesian team members (configurations 2-0 and 4-0).
\end{itemize}

\begin{figure}
 \centering
 \includegraphics[width=\linewidth]{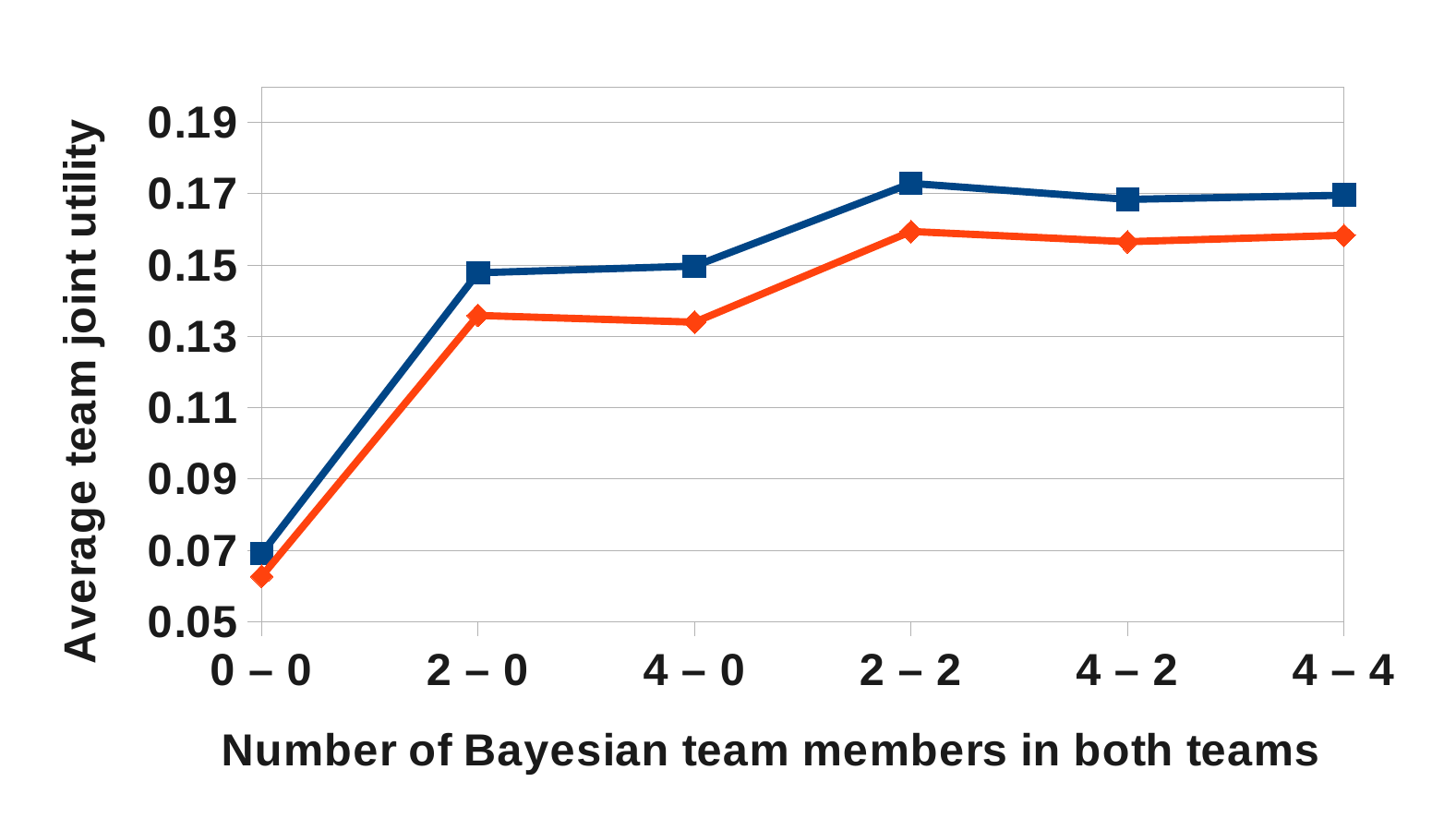}

\caption{Evolution of the average team joint utility with the different strategy profiles (Blue= group of travelers, Red= board of managers). }
\label{fig:teamvsteam}
\end{figure}

We configure the parameters as we did in our previous experiment. The results of the experiment can be observed in Figure \ref{fig:teamvsteam}. The blue points represent the average team joint utility for the group of travelers, whereas the red points represent the average team joint utility for the board of managers. The graphic shows an increasing average team joint utility as the total of Bayesian team members in both sides increases. The worst results for both teams are obtained when all of the team members act as the basic team member (0-0). This result is explainable due to the large number of negotiations that finished with no agreement (80 out of 160 negotiations, a 50\% of failure). As long as one of the sides implements the Bayesian strategy (2-0, 4-0), both teams benefit by obtaining higher average team joint utilities. An ANOVA test with Bonferroni post-hoc analysis ($\alpha=0.05$) confirmed that both configurations 2-0 and 4-0 obtain statistically different and higher team joint utilities for both teams than the configuration 0-0. This result confirmed our initial hypothesis H4.

It can be observed that the next relevant increase in the average team joint utility of both teams is present as long as both sides apply the Bayesian strategy (2-2, 4-2, 4-4). An ANOVA test with Bonferroni post-hoc analysis ($\alpha=0.05$) revealed that the averages obtained by configurations 2-2, 4-2, and 4-4 are statistically different and higher than the averages obtained by 2-0 and 4-0. Hence, both teams obtain higher average team joint utilities as long as both teams include Bayesian team members, supporting our hypothesis H5.

In conclusion, we have been able to determine that when two teams face each other by means of the proposed model, both teams benefit by obtaining higher team joint utilities when Bayesian team members participate in the negotiation. This is especially true when Bayesian team members are distributed between both teams, which obtains the highest team joint utilities for both sides.

\subsection{Analyzing the impact of Bayesian weights for the proposal of unpredictable partial offers}
\label{exp-2}

Recalling from Section \ref{Sec-BayesianTeamMember}, there are two weight parameters that control how important the opponent's and team's preferences are while generating the unpredictable partial offer (respectively $w_{op}, w_{A}$). $w_{A}$ represents how important it is for the team members to make an unpredictable partial offer that is acceptable for the team, whereas $w_{op}$ represents how important it is for us to make an unpredictable partial offer that is acceptable for the opponent. The use of these weights is not trivial, since one should consider that, it only refers to the acceptability of the unpredictable partial offer by one of the two parties. A complete offer is composed by the predictable and unpredictable issues. Therefore, for instance, using a high value of $w_{op}$ may not have the desired effect on the opponent unless unpredictable issues are important for the opponent. Additionally, one should also consider that the more acceptable an unpredictable is for the team/opponent, the more utility it should report.

In this experiment we explore the impact of these weights in a wide variety of situations. More specifically,  we study how different values for $w_{op}$  and $w_{A}$ affect situations where the team gives more importance to unpredictable issues than the opponent, situations where the opponent gives more importance to unpredictable issues than the team, and situations where both team and the opponent give the same importance to unpredictable issues.

To assess the importance given by an agent to unpredictable partial issues, we consider the sum of unpredictable issue weights in its utility function.
\begin{equation}
 I_{a}=\underset{j \in \mbox{\textit{UN}}}{\sum}w_{i,j}
\end{equation}
We consider that when $I_{a} \in [0.0, 0.33]$ the agent $a$ gives low importance to unpredictable issues, when $I_{a} \in \;\;[0.33, 0.66]$ it gives average importance to unpredictable issues, and when $I_{a} \in \;\;[0.66, 1.0]$ the agent gives high importance to unpredictable issues. We generated $8$ random negotiation cases where team members give a high importance to unpredictable issues and the opponent gives low ($4$ cases) and average ($4$ cases) importance to unpredictable issues, $8$ different randomly generated negotiation cases where team members give a low importance to unpredictable issues and the opponent gives average ($4$ cases) and high ($4$ cases) importance to unpredictable issues, and $12$ negotiation cases where the opponent and the team give the same importance to unpredictable issues (4 cases where both give low importance, 4 cases where both give average importance, and 4 cases where both give high importance).

We tested three configurations for Bayesian teams: standard Bayesian members that give the same importance to the acceptability of the unpredictable partial offer by the opponent and the team $w_{A}=w_{op}=0.5$ (Normal), Bayesian members that give more importance to the acceptability of the unpredictable partial offer by the opponent $w_{A}=0.25 \;\; w_{op}=0.75$ (Opponent Oriented), and Bayesian members that give more importance to the acceptability of the unpredictable partial offer by the team $w_{A}=0.75 \;\; w_{op}=0.25$ (Team Oriented). As for the opponent's strategies, we selected Agent K, Nice TFT and Boulware.

In situations where the opponent gives more importance to unpredictable issues than the team, the team should be able to obtain higher average team joint utility by playing high values for $w_{op}$, satisfying opponent needs with unpredictable issues, and demanding more on predictable issues. If the team gives more importance to unpredictable issues, the team should be able to obtain higher average team joint utility by playing high values for $w_{A}$, satisfying opponent needs with predictable issues, and demanding more on unpredictable issues. If both teams give the same importance to unpredictable issues, the team should be able to guarantee higher average team joint utility by giving the same weight to $w_{op}$ and $w_{A}$ since no party gives more importance to unpredictable issues. Attending to these initial guesses, we formulated the following hypotheses:

\begin{itemize}
 \item \textbf{H6} When the opponent gives more importance to unpredictable issues, the team obtains higher team joint utility by using high values for $w_{op}$ and demanding more on predictable issues ($w_{A}=0.25 \;\; w_{op}=0.75$).
 \item \textbf{H7} When the team gives more importance to unpredictable issues, the team obtains higher team joint utility by using high values for $w_{A}$ and letting the opponent demand on predictable issues ($w_{A}=0.75 \;\; w_{op}=0.25$).
 \item \textbf{H8} When both parties give the same importance to unpredictable issues, the team obtains higher team joint utility by giving the same importance to the opponent's and team's preferences on unpredictable issues ($w_{A}=w_{op}=0.5$).
\end{itemize}

For each negotiation case, we repeated the negotiation 20 times in order to capture stochastic variations in strategies. Therefore, a total of $1680$ negotiations were carried out in this experiment. The results of this experiment can be observed in Table \ref{tab-weights}. It shows the average of the joint team utility and proportion of negotiations that finished with success (agreement) in the scenario.  An ANOVA test ($\alpha=0.05$) with Bonferroni post-hoc analysis was carried out to detect statistically different averages. The best configurations for each of the three scenarios are highlighted in bold font style. All of the claims of this experimental section are supported with the aforementioned ANOVA test with post-hoc analysis.

Focusing on the case of negotiations against Agent K, it can be appreciated that when unpredictable issues are more important for the opponent (bottom part of Table \ref{tab-weights}), the best results are obtained by taking an opponent oriented approach: proposing unpredictable partial offers that are likely to be acceptable for the opponent and satisfy remaining members' aspirations by demanding on predictable and compatible issues, which are less important for the opponent.
This finding supports H6. As for the scenario where unpredictable issues are more important for the team (middle part of Table \ref{tab-weights}), it is clearly observed that the best choice for team joint utility is to give a high weight to $w_{A}$, thus employing a team oriented approach. Since unpredictable issues are more important for the team, they should satisfy their needs as much as possible with proposed unpredictable partial offers and demand less on predictable issues, which are more important for the opponent, supporting our initial hypothesis H7.  Finally, the last scenario corresponds to the case where unpredictable issues have the same importance for both parties (top part of Table \ref{tab-weights}). In this case, there may be more conflict between the team and its opponent since the parties do not have a clear trade-off opportunity such as increasing the demand on unpredictable issue while decreasing the demand on predictable issues as appeared in two previous cases. One can observe that the best team joint utility is obtained when using the standard team members
($w_{A}=w_{op}=0.5$). Since both parties give the same importance to unpredictable issues, it seems natural to give the same importance to the acceptability of the unpredictable partial offer by the team and the opponent, supporting our hypothesis H8. The team oriented approach is clearly worse than the rest of approaches since many negotiations (only a 58.7\% were successful. See top part of Table \ref{tab-weights} ) ended in failure due to the team being too demanding and not satisfying the opponent's preferences.

When negotiating against Nice TFT and Boulware, the results are different. It can be appreciated that, generally, the team oriented approach always reports statistically better results from the point of view of the average team joint utility. This means that the team should indistinctly select highly acceptable unpredictable partial offers for the team when facing opponents like Nice TFT and Boulware agents. These findings do not support our hypotheses H6, H7, and H8, and drove a more in-depth analysis and study. There are two factors that should be taken into account. First, as we detected in Section \ref{Sec-exp1}, the Bayesian models that are learnt from both agents are not adequate due to the lack of learning samples detected in the case of Nice TFT and the misinterpretation of the importance of issue values in the case of the Boulware agent. This factor precludes the agents from finding good agreements for both parties when using an opponent oriented approach in scenarios where unpredictable issues are more important for the opponent. Since the Bayesian models misinterpret the preferences of Boulware and Nice TFT agents, it is not possible to create \textit{win-win} situations. Second, by selecting a team oriented approach, the Bayesian team members are selecting the more acceptable (i.e., the best) unpredictable partial offers for the team. These unpredictable partial offers report high utility for the team. Thus, once completed with predictable issues, these offers are expected to report high utility for the team. Differently to Agent K, Nice TFT and Boulware agents are not competitor agents since they will eventually accept the team's offer when the deadline is approaching. In the case of the Boulware agent, it  concedes quickly with respect to its aspirations as the deadline approaches, eventually meeting the requirements of the demanding team offer. As stated in \citep{baarslag13}, when the negotiation time is reaching its deadline, the Nice TFT agent will wait for an offer that is not expected to improve in the remaining time in order to secure an agreement. Hence, the team is able to exploit the other party by selecting the team oriented approach. Due to these circumstances, hypotheses H6, H7, and H8 were not supported for agents Boulware and Nice Tit-for-Tat.

In conclusion, hypotheses H6, H7 and H7 are only partially supported. In general, we have discovered that depending on the type of opponent agent, the values for weights $w_{op}$, $w_{A}$ have different effects on the negotiation. When facing exploitable agents like Nice TFT and Boulware, the team benefits if the team members take the team oriented approach, selecting those unpredictable partial offers that are more acceptable for the team. If the team faces competitors like Agent K, the team should match the values for $w_{op}$ and $w_{A}$ depending on the importance given by each party to unpredictable issues. It is acknowledged that, depending on the opponent and the desired behavior, the team should select different values for the Bayesian proposal weights. A mechanism for adjusting Bayesian weights based on the type of opponent is considered as future work.

 \begin{table*}
 \centering

  \begin{tabular}{c  c c  c c  c c}
\cline{1-7}
 \multicolumn{7}{c}{\textbf{Equal importance on unpredictable issues}}\\
 & \multicolumn{2}{c}{vs. Agent K} & \multicolumn{2}{c}{vs. Nice TFT} & \multicolumn{2}{c}{vs. Boulware} \\
   & T. Joint & \% Ag.  & T. Joint & \% Ag. & T. Joint & \% Ag. \\
 Normal &   \textbf{0.168} & 84.2  & 0.137 & 100 & \textbf{0.202} & 100   \\
 Opponent Oriented &  0.155 & 91.1  & 0.126 & 100  & 0.188 & 100   \\
 Team Oriented    & 0.116 & 58.7   & \textbf{0.188} & 100 &  \textbf{0.206} & 100   \\
  \end{tabular}

  \begin{tabular}{c  c c  c c  c c}
\cline{1-7}
 \multicolumn{7}{c}{\textbf{Unpredictable issues more important for the team}}\\
 & \multicolumn{2}{c}{vs. Agent K} & \multicolumn{2}{c}{vs. Nice TFT} & \multicolumn{2}{c}{vs. Boulware} \\
   & T. Joint & \% Ag.  & T. Joint & \% Ag. & T. Joint & \% Ag. \\
 Normal &  0.213 & 100    & 0.135 & 100  & 0.185 & 100     \\
 Opponent Oriented & 0.200 & 100   & 0.154 & 100  & 0.189 & 100      \\
 Team Oriented    & \textbf{0.248} & 100   & \textbf{0.196} & 100  & \textbf{0.213} & 100      \\
  \end{tabular}

 \centering
  \begin{tabular}{c  c c  c c  c c}
\cline{1-7}
 \multicolumn{7}{c}{\textbf{Unpredictable issues more important for the opponent}}\\
 & \multicolumn{2}{c}{vs. Agent K} & \multicolumn{2}{c}{vs. Nice TFT} & \multicolumn{2}{c}{vs. Boulware} \\
   & T. Joint & \% Ag.  & T. Joint & \% Ag. & T. Joint & \% Ag. \\
 Normal &  0.280 & 100   & 0.186 & 100  & 0.326 & 100     \\
 Opponent Oriented &  \textbf{0.296} & 100   & 0.192 & 100  & 0.300 & 100     \\
 Team Oriented    & 0.271 & 92.0  & \textbf{0.259} & 100  & \textbf{0.340} & 100     \\
\cline{1-7}
  \end{tabular}
 \caption{Impact of $w_{A}$ and $w_{op}$ on the average team joint utility in different scenarios and proportion of negotiations that finished with success (\% Ag.).}
\label{tab-weights}
 \end{table*}

\subsection{Analyzing the Impact of the Reservation Utility}
\label{exp-3}

In this experiment, we investigated the impact of the reservation utility of team members on the team joint utility. As explained in Section \ref{Sec-PreNego}, team members jointly prune a part of the negotiation space (i.e., a set of unpredictable partial offers) which does not contain, with absolute certainty, any unanimously acceptable offer. This pruning is related with the reservation utility of team members, which represents the minimum acceptable utility by team members. Any offer with a utility lower than the reservation utility is not acceptable for the agent.

Lower reservation utilities make it easier to obtain the needed utility by just setting compatible and predictable issues. Thus, each team member needs to prune less negotiation space with the unpredictable partial offers sent to the team mediator. Presumably, a joint list of forbidden unpredictable partial offers (i.e., the negotiation space that is pruned)  with lower reservation utilities is smaller than lists constructed with higher reservation utilities. This leaves more room for finding an agreement with the opponent. However, if team members have low reservation utilities, despite having more room for finding an agreement, they may end up with low utility agreements in the end. On the contrary, with higher reservation utilities, it is harder to obtain the needed utility with compatible and predictable issues. Therefore, each team member may need to prune more negotiation space and the joint list of forbidden unpredictable partial offers will be larger than the list constructed with lower reservation utilities. In fact, if team members set high aspirations with their reservation utility, it may end up with all the negotiation space being pruned. If an agreement is found under these conditions, it may lead to team members achieving high levels of utility.

\begin{table}
\small
 \centering
 \begin{tabular}{|l| l| l| l|}
\cline{1-4}
  & Similar & Avg. Similarity & Dissimilar \\

$RU_{a}=0.35$& 0.4\% & 11.6\% & 35.3\% \\
$RU_{a}=0.50$& 23.8\% & 34.2\% & 72.6\% \\
$RU_{a}=0.65$& 73.7\% & 81.8\% & 90.8\%\\
\cline{1-4}
 \end{tabular}
 \caption{Average percentage of unpredictable partial offers pruned in the pre-negotiation.}
 \label{res-4}
\end{table}

In this experiment, we test the impact of different levels of reservation utility on the team joint utility. More specifically, as we did in Section \ref{Sec-exp1}, we tested teams employing the Bayesian model against different families of strategies: competitor (i.e., Agent K), matcher (i.e., Nice Tit-for-Tat), inverter (i.e., Boulware), and conceder (i.e., conceder). As an additional dimension to our study, we also introduced preference similarity among team members. Therefore, teams are tested in the scenario where team members' preferences are dissimilar, the scenario where team members' preferences have an average degree of similarity, and the scenario where team members' preferences are similar.

We configured three types of Bayesian teams with different levels of reservation utilities:  a team with a relatively low reservation utility $RU_{a}=0.35$,  a team with a moderate reservation utility $RU_{a}=0.5$, and  a team with a high reservation utility $RU_{a}=0.65$. We expect that playing higher reservation utilities against competitor like Agent K will result in lower average team joint utility due to many negotiations ending with no agreement. On the other hand, we expected that playing high reservation utilities against Conceder agents, inverter agents like Boulware, and the special case of Nice Tit-for-Tat, would result in higher average team joint utility due to the fact that both agents should eventually concede towards the other parties' demands. In the case of Nice Tit-for-Tat, we have observed that when the deadline is approaching it attempts to secure a deal, making this agent candidate to be exploited by setting a high reservation utility. Therefore, we formulated the following hypotheses:

\begin{itemize}
 \item \textbf{H9} Playing a high reservation value for team members ($RU_{a}=0.65$) will result in the lowest average team joint utility against Agent K.
 \item \textbf{H10} Playing a high reservation value for team members ($RU_{a}=0.65$) will result in the highest average team joint utility against Conceder, Boulware, and Nice Tit-for-Tat.
\end{itemize}

These types of teams ($RU_{a}=0.35, RU_{a}=0.5, RU_{a}=0.65$) were faced in every scenario and negotiation case against every type of opponent for $20$ repetitions. We gathered information on the team joint utility and the utility obtained by the opponent, and an ANOVA ($\alpha=0.05$) with Bonferroni post-hoc analysis was carried out to determine results that are statistically better than the rest. All of the claims of this experimental section are supported by the ANOVA test with the Bonferroni post-hoc analysis.

\begin{table*}
 \centering
 \begin{tabular}{|l|l l|l l|l l|l l|}
\cline{1-9}
  \multicolumn{9}{|c|}{Similar} \\ \cline{1-9}
  &  \multicolumn{2}{c|}{Agent K} & \multicolumn{2}{c|}{Nice Tit-for-Tat} & \multicolumn{2}{c|}{Boulware} & \multicolumn{2}{c|}{Conceder} \\ \cline{1-9}
  &  T. Joint & \% Ag.  & T. Joint & \% Ag.  &  T. Joint & \% Ag.  &  T. Joint & \% Ag.  \\
$RU_{a}=0.35$  & 0.195 & 100       &  0.117 &  100       &    0.160 & 100             &  0.526 & 100               \\
$RU_{a}=0.50$&  0.263 & 100     & 0.164 & 100      &   0.224 & 100         &  0.557 & 100         \\
$RU_{a}=0.65$& \textbf{0.350} & 99.98 &   \textbf{0.286} & 100      &  \textbf{0.354} & 100        &   \textbf{0.635} & 100            \\
 \cline{1-9}
  \multicolumn{9}{|c|}{Average Similarity} \\ \cline{1-9}
  &  \multicolumn{2}{c|}{Agent K} & \multicolumn{2}{c|}{Nice Tit-for-Tat} & \multicolumn{2}{c|}{Boulware} & \multicolumn{2}{c}{Conceder} \\ \cline{1-9}
  &  T. Joint & \% Ag.  & T. Joint & \% Ag.  &  T. Joint & \% Ag.  &  T. Joint & \% Ag.  \\
$RU_{a}=0.35$  &  0.167 & 100             &   0.090 & 100     &   0.136 & 100                &   0.342 & 100              \\
$RU_{a}=0.50$&  \textbf{0.248} & 100     &  0.158 & 100     &   0.224 & 100            &   0.390 & 100           \\
$RU_{a}=0.65$& \textbf{0.242} & 74.2   &  \textbf{0.268}  & 100       &   \textbf{0.313} & 100        &  \textbf{0.470} & 100             \\
  \cline{1-9}
  \multicolumn{9}{|c|}{Dissimilar} \\ \cline{1-9}
  &  \multicolumn{2}{c|}{Agent K} & \multicolumn{2}{c|}{Nice Tit-for-Tat} & \multicolumn{2}{c|}{Boulware} & \multicolumn{2}{c|}{Conceder} \\ \cline{1-9}
  &  T. Joint & \% Ag.  & T. Joint & \% Ag.  &  T. Joint & \% Ag.  &  T. Joint & \% Ag.  \\
$RU_{a}=0.35$  & \textbf{0.193} & 100    &  0.115  & 100           &  0.173  & 100                 &  0.373    & 100             \\
$RU_{a}=0.50$&   \textbf{0.209} & 86.6    &  0.196   & 100     &   0.271   & 100          &    0.475 & 100             \\
$RU_{a}=0.65$& 0.068 & 18.3 &  \textbf{0.346} & 100        &   \textbf{0.409}   & 100             &   \textbf{0.580}  & 100             \\
\cline{1-9}
 \end{tabular}
 \caption{Average joint Utility (T. Joint)  for teams composed by Bayesian team members with different reservation utilities and proportion of negotiations that finished with success (\% Ag.).}
 \label{res-3}
\end{table*}


Table \ref{res-3} shows the results of this experiment in terms of the average joint utility.  A bold font style is used to highlight those Bayesian team configurations that are statistically the best option against each opponent. Additionally, the same table shows the proportion of negotiations that finished with success (agreement) in this experiment. Table \ref{res-4} shows the average percentage of unpredictable partial offers that were pruned in the pre-negotiation depending on the team configuration and team preference similarity.

With respect to H10, it can be observed that independently of the degree of dissimilarity among team members' preferences, team members obtained statistically better team joint utility by setting high reservation utilities ($RU_{a}=0.65$) against  Boulware, Conceder, and Nice Tit-for-Tat opponents. These results support our initial hypothesis H10.

 Nevertheless, H9 is only partially supported.  When facing Agent K, playing the highest reservation utility configuration lead to the highest team joint utilities when team members' preferences are similar or they have an average similarity (top and middle part of Table \ref{res-3}), which is opposite to our initial hypothesis H9. In this case, the demands of the team are still not high enough to preclude agreements with the competitor agent.  However, when team members' preferences are dissimilar (bottom part of Table \ref{res-3}), we can observe how setting a high reservation utility (i.e., $RU_{a}=0.65$) gradually becomes the worst possible course of action when facing Agent K, making H9 true for this situation. The reason for this behavior is mainly explained due to the decrease in the number of successful negotiations. If we observe  the similar scenario, the number of successful negotiations when facing Agent K and $RU_{a}=0.65$ is 99.98\%. If we change to average similarity scenarios, the number of successful negotiations is 74.2\%  when facing Agent K with $RU_{a}=0.65$. The same measure is decreased to 18.3\% in the dissimilar scenario. If we observe Table \ref{res-4}, as team dissimilarity increases, the number of unpredictable partial offers to be pruned is larger. This leaves less negotiation space to be played with Agent K. Agent K is a competitor agent that attempts to concede as less as possible by estimating the maximum utility that can be obtained from the opponent and employing a limit of compromise when the opponent takes a hard stance. First of all, if reservation utilities are high, it can be considered that team members play a hard stance. Second, if too much negotiation space is pruned, it may be feasible that the set of remaining unpredictable partial offers precludes Agent K from reaching its limit of compromise. Thus, employing high reservation utilities against a competitor agent like Agent K may result, as we have observed in this case, in an increase in the number of failed negotiations and lower team joint utilities.

In conclusion, H10 is supported by our findings, but H9 is only partially supported. We have observed that team members may benefit from playing high reservation utilities against Conceder, Boulware, and Nice Tit-for-Tat. If faced against competitors like Agent K, setting high reservation utilities may prune too much negotiation space, especially when team members are dissimilar. This results in negotiation spaces that may not contain the minimum limits established by competitor agents, thus, ending negotiations with failure. For other scenarios against agent K, the team can benefit from setting high reservation utilities since the resulting negotiation space still has room to accommodate an agreement with the competitor agent.  In general, team members should be cautious when setting the reservation utility since it may end up in more failures.

\subsection{Team performance with risk attitudes}



One scenario that should be considered in multi-agent systems is agent's attitude towards risk. Some team members may be more willing to choose actions that guarantee a safer agreement, while other may prefer to go for more profitable but less probable agreements. Classically, agents can show a risk seeking, a risk averse, and a risk neutral attitude. The goal of this experiment is determining how risk attitudes affect the performance of the proposed negotiation model. In this experiment, we decided to test three different patterns of behavior:

\begin{itemize}
 \item Risk averse team member: This team member selects from the list of available unpredictable partial offers (see Equation \ref{eq:partialof}, those that guarantee the current aspirations of the team member $s_{a}(t)$)  the best unpredictable partial offer according to the acceptance probability for the opponent by using the Bayesian mechanism proposed in this article. When the team sets predictable issue, the maximum utility obtainable with other unpredictable partial offers may be higher since they may provide a higher utility of oneself. However, in order to secure a deal, the team member selects the unpredictable partial offer that is supposed to be more acceptable to the opponent even if the maximum achievable utility is lower. Therefore, the team member bases its choices on the acceptability of the offer by the opponent party instead of the maximum achievable utility. 
  \item Risk seeking team member: This team member selects for the list of available unpredictable partial offers (see Equation \ref{eq:partialof}) the best unpredictable partial offer according to the utility reported by one's own utility function. Hence, this unpredictable partial offer represents the choice that enables the team member to get the maximum achievable utility in the negotiation domain. By filling predictable issues by the team, the team member is more likely to get closer to its maximum achievable utility, even if this action reduces acceptability of the proposed offer for the opponent. 
  \item Risk neutral team member: This behavior is represented by the Bayesian team member presented in this article.
\end{itemize}

Our initial hypothesis is that the proposed model is robust to risk attitudes. By robustness, we mean that the proposed negotiation model will be able to obtain a team joint utility higher than or comparable to the team joint utility obtained by other state-of-the-art models like SBV:
\begin{itemize}
 \item \textbf{H11} Different configurations of team members' risk attitudes will yield a team joint utility that is statistically higher or equal to the results obtained by SBV.
\end{itemize}

The experimental parameters of the previous experiments were repeated ($RU_{a}=0.5$, $\beta_{a}=U[0.5,1]$), and we selected four different team configurations. The first one is composed by four neutral team members (Neutral), the second is composed by four risk seeking team members (Seeker), the third is composed by three risk averse team members (Averse), and the last team configuration is composed by two neutral team members, one risk seeker, and one risk averse (Mix).  As in the previous experiments, the team members faced different opponent profiles in scenarios where team members' preference profiles are similar, scenarios with average similarity of team members' preferences, and scenarios where team members have dissimilar preferences. The team faces the same opponent strategies presented in the previous experiments. We gathered information on the team joint utility.

 \begin{table}
 \small
  \centering
  \begin{tabular}{|l|l|l|l|l|}
 \cline{1-5}
   \multicolumn{5}{|c|}{Similar} \\ \cline{1-5}
   &  K & N. TFT & B. & C. \\ \cline{1-5}
 Averse  &  \textbf{0.241}   &  0.160   &    0.238      &  0.372      \\
 Neutral & \textbf{0.259}   &  0.164 &  0.224   &  0.557  \\
 Mix &  \textbf{0.250}  &  0.216 &    0.267  &  0.520   \\
 Seeker & 0.169 &   \textbf{0.248} & \textbf{0.283} & \textbf{0.615} \\
 SBV &  0.181        &      0.150       &     0.184      &  0.552     \\
  \cline{1-5}
   \multicolumn{5}{|c|}{Average Similarity} \\ \cline{1-5}
    &  K & N. TFT & B. & C. \\ \cline{1-5}
 Averse  &   0.188  & 0.147    &    0.186      &  0.308      \\
 Neutral &  \textbf{0.248}  &  \textbf{0.158}  &  \textbf{0.224}    &  0.390   \\
 Mix &  \textbf{0.231} &  \textbf{0.171} &  \textbf{0.222}   &  0.383   \\
 Seeker & 0.149 & \textbf{0.162} & \textbf{0.224} & \textbf{0.477} \\
 SBV &   0.168       &   0.137          &    0.170           &  0.324     \\
   \cline{1-5}
   \multicolumn{5}{|c|}{Dissimilar} \\ \cline{1-5}
    &  K & N. TFT & B. & C. \\ \cline{1-5}
 Averse  &  0.203   &  0.167  &     0.245     &    0.373    \\
 Neutral &  0.209  &  0.196  &  0.271   &  0.475   \\
 Mix & \textbf{0.228}  &  0.183  &  0.265   &  0.452   \\
 Seeker & 0.060 & \textbf{0.247} & \textbf{0.294} & \textbf{0.566} \\
 SBV &      0.07    &      0.16       &     0.128          &  0.257     \\
 \cline{1-5}
  \end{tabular}
  \caption{Team joint utility obtained by the different team risk configurations. K (Agent K), N. TFT (Nice Tit-for-Tat), B. (Boulware), C. (Conceder)}
  \label{res-het}
 \end{table}

 The results of this experiment can be found on Table \ref{res-het}. Results highlighted in bold represent the statistically higher team joint utility configurations for each negotiation scenario (ANOVA, $\alpha=0.05$). As it can be observed, the risk seeking configuration usually gets a higher team joint utility as long as the opponent faced is not a competitor. This configuration is able to obtain one of the best team joint in all negotiation scenarios with Nice Tit-for-Tat, Boulware, and Conceder. When negotiating against Agent K , many negotiations end with failure since the opponent also has high aspirations. Other opponents like Nice Tit-for-Tat, Boulware, or Conceder match the high aspirations of the risk seeking team at one point or another of the negotiation, resulting in a higher team joint utility.

 The mixed configuration usually gets the next highest team joint utility, being able to obtain statistically equivalent results in some scenarios like negotiating against Nice Tit-for-Tat in the average similarity scenario, and statistically higher results like negotiations against Agent K in dissimilar scenarios. This configuration obtains the best team joint utility in all of the scenarios involving against Agent K.

 The neutral configuration, obtains statistically good results in some scenarios where team members have similar preferences and the Bayesian mechanism is able to learn the opponent preferences properly (e.g., negotiating against Agent K, some scenarios against Boulware and Nice Tit-for-Tat). The risk averse configuration is only able to obtain some of the top results in very specific scenarios like negotiations against in similar settings against Agent K.

 However, in no case any of the team configurations proposed in this article is worse than SBV. All of the results are statistically higher than those obtained by SBV, except for those obtained by the risk seeking configuration against Agent K, which are statistically equivalent to those obtained by SBV. Therefore, despite being affected by team heterogeneity, the proposed model is robust and it is able to obtain results that are at least equal to the state-of-the-art (i.e., SBV), outperforming it in many situations. This result supports our initial hypothesis H11.

\section{Related Work}
\label{Sec-Discussion}

The contributions of this article to the automated negotiation community can be divided into two different categories: general contributions to the field of automated negotiation, and contributions to the specific field of agent-based negotiation teams. Next, we analyze the contributions to each of these fields.

\subsection{Automated negotiation with single individual parties}

The artificial intelligence community has focused on bilateral or multi-party negotiations where parties are composed of single individuals. The main contribution of our work with respect to the general field of automated negotiation is that we support negotiation parties composed by multiple individuals. Apart from that, in the next paragraphs we discuss other contributions of our present negotiation framework with respect to works in automated negotiation.

Faratin \emph{et al.} \citep{faratin98} introduced some of the most widely used families of concession tactics in negotiation. The authors proposed concession strategies for negotiation issues that are a mix of different families of concession tactics. The authors divide these concession tactics into three different families: (i) time-dependent concession tactics; (ii) behavior-dependent concession tactics; and (iii) resource-dependent tactics. Our negotiation framework also considers time as crucial element in negotiation. Therefore, team members employ time tactics inspired in those introduced by Faratin \emph{et al.} However, the authors do not propose any explicit preference learning mechanism.

In Zeng and Sycara \citep{zeng98}, the authors argue about the benefits of using Bayesian models in negotiation and they study a bilateral negotiation case where the buyer attempts to learn the reservation price of the seller by updating its beliefs with Bayesian learning. Despite the fact that it introduces the use of Bayesian learning in negotiation, the article only focuses on single issue models. One of our team member models also uses Bayesian learning as a method for learning other agents' preferences. The main different resides in the fact that our Bayesian approach attempts to model which instantiations of unpredictable issues are acceptable for the opponent and the team in multi-issue negotiations.

Ehtamo et al. \citep{ehtamo01} propose a \textit{mediated} multi-party negotiation protocol which looks for joint gains in an iterated way. The algorithm starts from a tentative agreement and moves in a direction according to what the agents prefer regarding some offers' comparison. Results showed that the algorithm converges quickly to Pareto optimal points. However, the work proposed by Ehtamo et al. does not support unpredictable issues and multiple individual parties.

Klein et al. \citep{klein03} propose a \textit{mediated} negotiation model which can be extended to multiple parties. Their main goal is to provide solutions for negotiation processes that use complex utility functions to model agents' preferences. The negotiation attributes are binary and no longer independent. Our work supports multiple individual parties and negotiation issues with unrestricted domain (e.g., real, integer, discrete, binary, etc.).

In Coehoorn \emph{et al.} \citep{coehoorn04}, the authors propose the use of kernel density estimation for the estimation of the importance weights of the linear additive utility function. The agent calculates tuples composed of the difference between pairs of consecutive offers, the estimated weight for the issue, and the probability density of the weight. These tuples form a three dimensional kernel that is used along the other kernels to calculate an estimation of the real issue weight. Our proposed model is capable of learning in negotiations where domains are also composed by discrete issues, which is not supported by Coehoorn \emph{et al.}  Moreover, the learning mechanism proposed for our team members deals with the information in one single negotiation, whereas the aforementioned mechanism learns over several negotiations.

Later, Narayanan \emph{et al.}\citep{narayanan06} present a negotiation framework where pairs of agents negotiate over a single issue. The authors assume that agents' strategies may change over time. Non-stationary Markov chains and Bayesian learning are employed to tackle the uncertainty in this domain, and eventually converge towards the optimal negotiation strategy. In our case, we focus on one single negotiation process, and our team members learn over the information provided by the current negotiation. Additionally, we consider negotiations where multiple issues are involved.

Another example of the use of Bayesian learning in negotiation is presented by Buffett \emph{et al.} \citep{buffett07}. In the aforementioned article, a bilateral framework is presented in a domain where agents negotiate over a set of binary issues.  A Bayesian classifier is employed to classify opponent's preferences into classes of preference relations. Groups of similar preference relations are grouped according to the k-means algorithm prior to the negotiation process. Our model does not require learning prior to the negotiation, which may require a costly learning process every negotiation to avoid domain dependent classifiers. Moreover, we consider any kind of issue type in the negotiation domain.

Carbonneau et al. \citep{carbonneau08} propose a neural network that takes as input the negotiation history of a bilateral negotiation with continuous issues and an offer to make an estimation of the opponent's counter-offer. This approach requires that an artificial neural network is trained per negotiation case. Similarly, the same authors propose an improvement over their previous work in \citep{carbonneau11}. It aims to make a predictive model that does not depend on the negotiation case. The model takes pairs of negotiation issues as inputs of the neural network, where one of the issues is considered the primary issue (i.e., independent variable) and the other issue is considered the secondary issue (i.e., dependent variable). Differently to these works, our proposed model does not rely on information from past negotiations. It only employs information gathered in the present process.

Robu et al.\citep{robu08} introduce a bilateral negotiation model where agents represent their preferences by means of utility graphs. The negotiation domain is formed of bundles of items that can be either included or excluded in a final deal. Utility graphs are graphical models that relate negotiation issues that are dependent. Nodes represent negotiation issues whereas arcs connect issues that have some joint effect on the utility function (i.e., positive for complementary issues, and negative for substitutable issues). Hence, utility graphs represent binary dependencies between issues. The authors propose a negotiation scenario where the buyer's preferences and the seller's preferences are modeled through utility graphs. The seller is the agent that carries out a more thorough exploration of the negotiation space in order to search for agreements where both parties are satisfied. With this purpose, the seller builds a model of the buyer's preferences based on historic information of past deals and expert knowledge about the negotiation domain. Our work does not consider dependencies between issues, however it is capable of supporting every type of domain for negotiation issues (i.e., real, binary, discrete, etc.).

In \citep{aydogan12}, a concept-based learning method is proposed for modeling opponent preferences and generating well-targeted offers. In that method, the preferences of the opponent are represented via disjunctive and conjunctive constraints. In this article, our aim is also to find the agreement earlier by means of learning the other participants' preferences but the preferences are represented by means of additive utility functions. In our case, an offer that is rejected by the opponent may become acceptable over time because the opponent may concede; therefore, we choose a probabilistic learning method.

Williams et al. \citep{williams12} present a negotiation framework for coordinating multiple bilateral negotiations with different opponents. The agent simultaneously negotiates with different opponent in order to acquire a desired good at the best possible condition. The framework makes use of optimization techniques and probabilistic information in order to carry out this coordination. Similarly, Mansour et al. \citep{mansour12} present a meta-strategy for coordination multiple negotiations with different sellers. The meta-strategy adjusts the concession speed according to the current state of the multiple negotiation threads. The problem of agent-based negotiation teams is different since multiple agents collaborate in the same party to get a deal from an opponent, instead of competing between each other to get an individual deal.

\subsection{Agent-based negotiation teams}

As far as we are concerned,  only our previous works \citep{sanchez-anguix11,sanchez-anguix12,sanchez-anguix12b} have considered negotiation teams in computational models. In \citep{sanchez-anguix11,sanchez-anguix12} four different computational models for a negotiation team negotiating with a single opponent are presented. These four models attempt to gather four different minimum levels of unanimity regarding team decisions: representative approach (RE, no unanimity), similarity simple voting (SSV, majority/plurality), similarity Borda voting (SBV, semi-unanimity), and full unanimity mediated (FUM, unanimity).

The RE model is based on the selection of team members as representative of the team. The representative acts on behalf of the group by taking decisions according to its own negotiation strategy and utility function.

SSV and SBV are models based on the presence of a mediator that coordinates voting processes. In the case of SSV, a majority voting is employed to determine whether or not the opponent's offer is accepted, and a majority voting is used to select which offer is sent from those offers proposed by team members to be sent to the opponent. In the case of SBV, a unanimity voting is designated as the mechanism to decide if opponent's offers are accepted and Borda count is used to decide which offer proposed by team members is sent to the opponent. In both cases, SSV and SBV team members decide which offers are sent to the opponent using similarity heuristics that consider the opponent's and team's last offers.

FUM is a mediated model where the opponent's offers are evaluated by means of a unanimity voting process, and team's offers are built issue per issue by aggregation rules. These models were introduced at AAMAS 2011 \citep{sanchez-anguix11} as the first approach to tackle problems involving negotiation teams.

Later, we studied the special properties of FUM, given that it was the intra-tem negotiation model capable of guaranteeing unanimity regarding team decisions at each negotiation round \citep{sanchez-anguix12b}. We proved how unanimity is guaranteed in FUM, how the intra-team strategy us robust against certain types of manipulation attacks, and how team members did not have incentive to strongly deviate from the proposed model.

We provided a full fledged experimental analysis of the four intra-team strategies in different negotiation environments \citep{sanchez-anguix12}. The results showed that FUM was able to guarantee better results for the negotiation team in most negotiation scenarios. Only SBV is able to guarantee similar results in a limited number of negotiation cases.

Even though these four models cover different levels of unanimity with regards to team decisions, they were initially designed to provide solutions for negotiation domains that are exclusively composed by \textit{predictable and compatible} issues among team members (e.g., price, quality, and dispatch time in a team of buyers).

Domains exclusively composed by \textit{compatible and predictable} issues among team members cover a range of feasible negotiation domains. However, a relatively large number of negotiation domains also include \textit{unpredictable} issues among team members. RE, SSV, and SBV can be easily adapted to domains that include \textit{unpredictable} issues among team members by using a different similarity heuristic. FUM, the model that guarantees unanimity with regards to team decisions, cannot be used in these domains since, in the offer proposal mechanism, it aggregates issue values based on the fact that all of the negotiation issues are \textit{compatible and predictable} among team members.

 As stated, whenever it is possible, it is desirable for the final agreement to be unanimously acceptable for all of the team members since the opposite situation may end up in users perceiving unfairness, which may affect commitment to the decision, group attachment, and trust \citep{korsgaard95}, and  users that are not satisfied with agreements found automatically may end up leaving the electronic application.

 The model proposed in this article advances the stat of the art in agent-based negotiation teams by solving both problems:  it guarantees  unanimity with regards to team decisions, and it supports \textit{unpredictable} negotiation issues, which were not specifically supported in previous models. For that matter, the negotiation domains is split between those issues that are \textit{compatible and predictable} among team members, and those issues that are \textit{unpredictable}.  In the former case, part of the mechanism employed in FUM is employed. By doing so, the model is capable of guaranteeing that team members are able to get as much as they need from \textit{predictable and compatible} issues. In the latter case,  the team discards those combinations of unpredictable partial offers that preclude the team from reaching unanimously acceptable agreements, even if the most is obtained from \textit{predictable and compatible} issues. Then, in the offer proposal mechanism team members select one of remaining unpredictable partial offers, and complete the \textit{predictable and compatible} issue with the values that they need to make it at least unanimously acceptable. The idea behind this splitting, is cooperating as much as possible on those issues that are \textit{predictable and compatible} among team members to create less intra-team conflict in \textit{unpredictable} issues.

\section{Conclusions \& Future Work}

In this article we have presented a new mediated team negotiation model for a team as a multi-individual party negotiating with an opponent in the alternating offers protocol. The present model is capable of assuring unanimously acceptable agreements for all of the team members. It takes advantage of the categorization of negotiation issues as predictable and compatible, and unpredictable. The former are those issues whose preferential order over issue values is known from the negotiation domain and it is common among team members (e.g., price in a team of buyers), whereas the latter are those issues whose preferential order over issues values is not known in the negotiation domain. In the case of predictable and compatible issues, there is full potential for cooperation among team members since if one of the team members demands more from the issue, the other team members are also benefited. Our negotiation model takes advantage of this property. During the pre-negotiation, each team member shares with a team mediator those unpredictable partial offers (i.e., partial offers that have all of the unpredictable issues instantiated) that, even if the team demands the most from predictable issues, preclude the agent from achieving its reservation value. A joint list forbidden unpredictable partial offers is constructed by the team mediator from the lists received from the team members. In the negotiation, the team mediator coordinates a unanimity voting process to decide whether or not to accept offers received from the opponent. As for the mechanism employed to decide on which offer should be sent to the opponent, the team mediator coordinates two processes: a proposing and voting process where each team member suggests an unpredictable partial offer not included in the forbidden list followed by a Borda voting on candidates received, and an iterated process where predictable issues are set issue per issue attending to the demands of the team members.

We have proposed two different types of team members for the current model: a basic team member that proposes unpredictable partial offers during the negotiation solely guided by its own utility function, and a Bayesian team member that suggests unpredictable partial offers based on the preferences of the team and the preferences of the opponent. Results have shown that, as long as preferential conflict is present in the team, team members have an incentive to employ the Bayesian strategy over the basic strategy. In any case, we have shown that both approaches outperform other existing models for negotiation teams. Moreover, we have determined that when two negotiation teams face each other, both teams benefit from including Bayesian team members in the negotiation. Finally, we have shown that team members may benefit from playing higher reservation utilities against conceders, matchers and inverters. Nevertheless, setting high reservation utilities may become the worst option as team members' preferences are more dissimilar and the opponent plays a competitor strategy.

The topic of agent-based negotiation teams remains largely unexplored compared to negotiation involving single individual parties. The present work has focused on agent-based negotiation teams where team members have different preferences but they share the same knowledge regarding the negotiation domain. One potential area of research is modeling negotiation teams where team members differ in their knowledge of the negotiation domain and their skills related to the negotiation process. As another line of future work, one could consider the problem of forming negotiation teams based on the individual list of unpredictable partial offers. Lists that are more similar may suggest team members that are more similar, which, if put together in the same team, may result in more cooperative teams. Related to team formation, dynamic negotiation teams where team members may join and leave the team during the negotiation can be considered an alternative line of research.

\section*{Acknowledgements}
This research is partially supported by TIN2012-36586-C03-01 of the Spanish government and PROMETEOII/2013/019 of Generalitat Valenciana. Other part of this research is supported by the Dutch Technology Foundation STW, applied science division of NWO and the Technology Program of the Ministry of Economic Affairs; the Pocket Negotiator project with grant number VICI-project 08075.

\bibliographystyle{elsarticle-num-names}
\bibliography{mybib}

\begin{thebibliography}{62}
\providecommand{\natexlab}[1]{#1}
\providecommand{\url}[1]{\texttt{#1}}
\providecommand{\urlprefix}{URL }
\expandafter\ifx\csname urlstyle\endcsname\relax
  \providecommand{\doi}[1]{doi:\discretionary{}{}{}#1}\else
  \providecommand{\doi}[1]{doi:\discretionary{}{}{}\begingroup
  \urlstyle{rm}\url{#1}\endgroup}\fi
\providecommand{\bibinfo}[2]{#2}

\bibitem[{Ngai and Wat(2002)}]{ngai02}
\bibinfo{author}{E.~W. Ngai}, \bibinfo{author}{F.~Wat}, \bibinfo{title}{A
  literature review and classification of electronic commerce research},
  \bibinfo{journal}{Information \& Management}
  \bibinfo{volume}{39}~(\bibinfo{number}{5}) (\bibinfo{year}{2002})
  \bibinfo{pages}{415--429}.

\bibitem[{Grieger(2003)}]{grieger03}
\bibinfo{author}{M.~Grieger}, \bibinfo{title}{Electronic marketplaces: A
  literature review and a call for supply chain management research},
  \bibinfo{journal}{European journal of operational research}
  \bibinfo{volume}{144}~(\bibinfo{number}{2}) (\bibinfo{year}{2003})
  \bibinfo{pages}{280--294}.

\bibitem[{Wareham et~al.(2005)Wareham, Zheng, and Straub}]{wareham05}
\bibinfo{author}{J.~Wareham}, \bibinfo{author}{J.~G. Zheng},
  \bibinfo{author}{D.~Straub}, \bibinfo{title}{Critical themes in electronic
  commerce research: a meta-analysis}, \bibinfo{journal}{Journal of Information
  Technology} \bibinfo{volume}{20}~(\bibinfo{number}{1}) (\bibinfo{year}{2005})
  \bibinfo{pages}{1--19}.

\bibitem[{Guttman et~al.(1998)Guttman, Moukas, and Maes}]{guttman98}
\bibinfo{author}{R.~H. Guttman}, \bibinfo{author}{A.~G. Moukas},
  \bibinfo{author}{P.~Maes}, \bibinfo{title}{Agent-mediated electronic
  commerce: a survey}, \bibinfo{journal}{Knowledge Engineering Review}
  \bibinfo{volume}{13}~(\bibinfo{number}{2}) (\bibinfo{year}{1998})
  \bibinfo{pages}{147--159}, ISSN \bibinfo{issn}{0269-8889},
  \doi{\bibinfo{doi}{http://dx.doi.org/10.1017/S0269888998002082}}.

\bibitem[{Sierra and Dignum(2001)}]{sierra01}
\bibinfo{author}{C.~Sierra}, \bibinfo{author}{F.~Dignum},
  \bibinfo{title}{Agent-mediated electronic commerce: Scientific and
  technological roadmap}, in: \bibinfo{booktitle}{Agent Mediated Electronic
  Commerce}, \bibinfo{publisher}{Springer}, \bibinfo{pages}{1--18},
  \bibinfo{year}{2001}.

\bibitem[{Oliveira and Rocha(2001)}]{oliveira01}
\bibinfo{author}{E.~Oliveira}, \bibinfo{author}{A.~Rocha},
  \bibinfo{title}{Agents advanced features for negotiation in electronic
  commerce and virtual organisations formation process}, in:
  \bibinfo{booktitle}{Agent Mediated Electronic Commerce},
  \bibinfo{publisher}{Springer}, \bibinfo{pages}{78--97}, \bibinfo{year}{2001}.

\bibitem[{He et~al.(2003)He, Jennings, and Leung}]{he03}
\bibinfo{author}{M.~He}, \bibinfo{author}{N.~R. Jennings},
  \bibinfo{author}{H.-F. Leung}, \bibinfo{title}{On agent-mediated electronic
  commerce}, \bibinfo{journal}{Knowledge and Data Engineering, IEEE
  Transactions on} \bibinfo{volume}{15}~(\bibinfo{number}{4})
  (\bibinfo{year}{2003}) \bibinfo{pages}{985--1003}.

\bibitem[{Lomuscio et~al.(2003)Lomuscio, Wooldridge, and Jennings}]{lomuscio03}
\bibinfo{author}{A.~R. Lomuscio}, \bibinfo{author}{M.~Wooldridge},
  \bibinfo{author}{N.~R. Jennings}, \bibinfo{title}{A Classification Scheme for
  Negotiation in Electronic Commerce}, \bibinfo{journal}{Group Decision and
  Negotiation} \bibinfo{volume}{12} (\bibinfo{year}{2003})
  \bibinfo{pages}{31--56}, ISSN \bibinfo{issn}{0926-2644}.

\bibitem[{Nguyen and Jennings(2005)}]{nguyen05}
\bibinfo{author}{T.~D. Nguyen}, \bibinfo{author}{N.~R. Jennings},
  \bibinfo{title}{Managing commitments in multiple concurrent negotiations},
  \bibinfo{journal}{Electronic Commerce Research and Applications}
  \bibinfo{volume}{4}~(\bibinfo{number}{4}) (\bibinfo{year}{2005})
  \bibinfo{pages}{362 -- 376}, ISSN \bibinfo{issn}{1567-4223},
  \bibinfo{note}{developments in intelligent support for e-Commerce negotiation
  applications The 6th International Conference on Electronic Commerce}.

\bibitem[{Buffett and Spencer(2007)}]{buffett07}
\bibinfo{author}{S.~Buffett}, \bibinfo{author}{B.~Spencer}, \bibinfo{title}{A
  bayesian classifier for learning opponents' preferences in multi-object
  automated negotiation}, \bibinfo{journal}{Electronic Commerce Research and
  Applications} \bibinfo{volume}{6}~(\bibinfo{number}{3})
  (\bibinfo{year}{2007}) \bibinfo{pages}{274--284}.

\bibitem[{lau(2007)}]{lau07}
\bibinfo{title}{Towards a web services and intelligent agents-based negotiation
  system for \{B2B\} eCommerce}, \bibinfo{journal}{Electronic Commerce Research
  and Applications} \bibinfo{volume}{6}~(\bibinfo{number}{3})
  (\bibinfo{year}{2007}) \bibinfo{pages}{260 -- 273}, ISSN
  \bibinfo{issn}{1567-4223}.

\bibitem[{Chan et~al.(2008)Chan, Cheng, and Hsu}]{chan08}
\bibinfo{author}{C.-C.~H. Chan}, \bibinfo{author}{C.-B. Cheng},
  \bibinfo{author}{C.-H. Hsu}, \bibinfo{title}{Bargaining strategy formulation
  with CRM for an e-commerce agent}, \bibinfo{journal}{Electronic Commerce
  Research and Applications} \bibinfo{volume}{6}~(\bibinfo{number}{4})
  (\bibinfo{year}{2008}) \bibinfo{pages}{490--498}.

\bibitem[{Klein et~al.(2003{\natexlab{a}})Klein, Faratin, Sayama, and
  Bar-Yam}]{klein03}
\bibinfo{author}{M.~Klein}, \bibinfo{author}{P.~Faratin},
  \bibinfo{author}{H.~Sayama}, \bibinfo{author}{Y.~Bar-Yam},
  \bibinfo{title}{Protocols for Negotiating Complex Contracts},
  \bibinfo{journal}{IEEE Intelligent Systems}
  \bibinfo{volume}{18}~(\bibinfo{number}{6})
  (\bibinfo{year}{2003}{\natexlab{a}}) \bibinfo{pages}{32--38}, ISSN
  \bibinfo{issn}{1541-1672}.

\bibitem[{Williams et~al.(2011)Williams, Robu, Gerding, and
  Jennings}]{williams11}
\bibinfo{author}{C.~Williams}, \bibinfo{author}{V.~Robu},
  \bibinfo{author}{E.~Gerding}, \bibinfo{author}{N.~Jennings},
  \bibinfo{title}{Using gaussian processes to optimise concession in complex
  negotiations against unknown opponents}, in: \bibinfo{booktitle}{Proceedings
  of the 22nd International Joint Conference on Artificial Intelligence
  (IJCAI'11)}, \bibinfo{pages}{432--438}, \bibinfo{year}{2011}.

\bibitem[{Baarslag et~al.(2012)Baarslag, Hindriks, Jonker, Kraus, and
  Lin}]{baarslag12}
\bibinfo{author}{T.~Baarslag}, \bibinfo{author}{K.~Hindriks},
  \bibinfo{author}{C.~Jonker}, \bibinfo{author}{S.~Kraus},
  \bibinfo{author}{R.~Lin}, \bibinfo{title}{The First Automated Negotiating
  Agents Competition (ANAC 2010)}, in: \bibinfo{booktitle}{New Trends in
  Agent-Based Complex Automated Negotiations}, vol. \bibinfo{volume}{383} of
  \emph{\bibinfo{series}{Studies in Computational Intelligence}},
  \bibinfo{publisher}{Springer Berlin / Heidelberg}, ISBN
  \bibinfo{isbn}{978-3-642-24695-1}, \bibinfo{pages}{113--135},
  \bibinfo{year}{2012}.

\bibitem[{Ball et~al.(2006)Ball, Coelho, and Vilares}]{ball06}
\bibinfo{author}{D.~Ball}, \bibinfo{author}{P.~S. Coelho},
  \bibinfo{author}{M.~J. Vilares}, \bibinfo{title}{Service personalization and
  loyalty}, \bibinfo{journal}{Journal of Services Marketing}
  \bibinfo{volume}{20}~(\bibinfo{number}{6}) (\bibinfo{year}{2006})
  \bibinfo{pages}{391--403}.

\bibitem[{Liang et~al.(2007)Liang, Lai, and Ku}]{liang07}
\bibinfo{author}{T.-P. Liang}, \bibinfo{author}{H.-J. Lai},
  \bibinfo{author}{Y.-C. Ku}, \bibinfo{title}{Personalized content
  recommendation and user satisfaction: Theoretical synthesis and empirical
  findings}, \bibinfo{journal}{Journal of Management Information Systems}
  \bibinfo{volume}{23}~(\bibinfo{number}{3}) (\bibinfo{year}{2007})
  \bibinfo{pages}{45--70}.

\bibitem[{Thompson(2003)}]{thompson03}
\bibinfo{author}{L.~Thompson}, \bibinfo{title}{The Mind and heart of the
  negotiator}, \bibinfo{publisher}{Prentice Hall Press},
  \bibinfo{address}{Upper Saddle River, NJ, USA}, \bibinfo{year}{2003}.

\bibitem[{Lai et~al.(2008)Lai, Sycara, and Li}]{lai08}
\bibinfo{author}{G.~Lai}, \bibinfo{author}{K.~Sycara}, \bibinfo{author}{C.~Li},
  \bibinfo{title}{A decentralized model for automated multi-attribute
  negotiations with incomplete information and general utility functions},
  \bibinfo{journal}{Multiagent and Grid Systems}
  \bibinfo{volume}{4}~(\bibinfo{number}{1}) (\bibinfo{year}{2008})
  \bibinfo{pages}{45--65}, ISSN \bibinfo{issn}{1574-1702}.

\bibitem[{Chevaleyre et~al.(2007)Chevaleyre, Endriss, Lang, and
  Maudet}]{chevaleyre07}
\bibinfo{author}{Y.~Chevaleyre}, \bibinfo{author}{U.~Endriss},
  \bibinfo{author}{J.~Lang}, \bibinfo{author}{N.~Maudet}, \bibinfo{title}{A
  Short Introduction to Computational Social Choice}, in:
  \bibinfo{booktitle}{SOFSEM 2007: Theory and Practice of Computer Science},
  vol. \bibinfo{volume}{4362} of \emph{\bibinfo{series}{Lecture Notes in
  Computer Science}}, \bibinfo{publisher}{Springer Berlin / Heidelberg},
  \bibinfo{pages}{51--69}, \bibinfo{year}{2007}.

\bibitem[{Taylor(2003)}]{taylor03}
\bibinfo{author}{H.~Taylor}, \bibinfo{title}{Most people are privacy
  pragmatists who, while concerned about privacy, will sometimes trade it off
  for other benefits}, \bibinfo{journal}{The Harris Poll} \bibinfo{volume}{17}
  (\bibinfo{year}{2003}) \bibinfo{pages}{19}.

\bibitem[{Faratin et~al.(1998)Faratin, Sierra, and Jennings}]{faratin98}
\bibinfo{author}{P.~Faratin}, \bibinfo{author}{C.~Sierra},
  \bibinfo{author}{N.~R. Jennings}, \bibinfo{title}{Negotiation Decision
  Functions for Autonomous Agents}, \bibinfo{journal}{International Journal of
  Robotics and Autonomous Systems} \bibinfo{volume}{24}~(\bibinfo{number}{3-4})
  (\bibinfo{year}{1998}) \bibinfo{pages}{159--182}.

\bibitem[{Zeng and Sycara(1998)}]{zeng98}
\bibinfo{author}{D.~Zeng}, \bibinfo{author}{K.~Sycara},
  \bibinfo{title}{Bayesian learning in negotiation},
  \bibinfo{journal}{International Journal of Human-Computer Studies}
  \bibinfo{volume}{48}~(\bibinfo{number}{1}) (\bibinfo{year}{1998})
  \bibinfo{pages}{125--141}, ISSN \bibinfo{issn}{1071-5819},
  \doi{\bibinfo{doi}{http://dx.doi.org/10.1006/ijhc.1997.0164}}.

\bibitem[{Klein et~al.(2003{\natexlab{b}})Klein, Faratin, Sayama, and
  Bar-Yam}]{klein03b}
\bibinfo{author}{M.~Klein}, \bibinfo{author}{P.~Faratin},
  \bibinfo{author}{H.~Sayama}, \bibinfo{author}{Y.~Bar-Yam},
  \bibinfo{title}{Negotiating Complex Contracts}, \bibinfo{journal}{Group
  Decision and Negotiation} \bibinfo{volume}{12}
  (\bibinfo{year}{2003}{\natexlab{b}}) \bibinfo{pages}{111--125}.

\bibitem[{Coehoorn and Jennings(2004)}]{coehoorn04}
\bibinfo{author}{R.~M. Coehoorn}, \bibinfo{author}{N.~R. Jennings},
  \bibinfo{title}{Learning on opponent's preferences to make effective
  multi-issue negotiation trade-offs}, in: \bibinfo{booktitle}{The 6th
  International Conference on Electronic Commerce (ICEC'04)},
  \bibinfo{pages}{59--68}, \bibinfo{year}{2004}.

\bibitem[{Sanchez-Anguix et~al.(2013)Sanchez-Anguix, Valero, Julian, Botti, and
  García-Fornes}]{sanchez-anguix11b}
\bibinfo{author}{V.~Sanchez-Anguix}, \bibinfo{author}{S.~Valero},
  \bibinfo{author}{V.~Julian}, \bibinfo{author}{V.~Botti},
  \bibinfo{author}{A.~García-Fornes}, \bibinfo{title}{{E}volutionary-aided
  negotiation model for bilateral bargaining in {A}mbient {I}ntelligence
  domains with complex utility functions}, \bibinfo{journal}{Information
  Sciences} \bibinfo{volume}{222} (\bibinfo{year}{2013})
  \bibinfo{pages}{25--46}.

\bibitem[{Aydo\u{g}an and Yolum(2012)}]{aydogan12}
\bibinfo{author}{R.~Aydo\u{g}an}, \bibinfo{author}{P.~Yolum},
  \bibinfo{title}{Learning opponent’s preferences for effective negotiation:
  an approach based on concept learning}, \bibinfo{journal}{Auton. Agents
  Multi-Agent Syst.} \bibinfo{volume}{24} (\bibinfo{year}{2012})
  \bibinfo{pages}{104--140}.

\bibitem[{Mannix(2005)}]{mannix05}
\bibinfo{author}{E.~Mannix}, \bibinfo{title}{Strength in Numbers: Negotiating
  as a Team}, \bibinfo{journal}{Negotiation}
  \bibinfo{volume}{8}~(\bibinfo{number}{5}) (\bibinfo{year}{2005})
  \bibinfo{pages}{3--5}.

\bibitem[{Halevy(2008)}]{halevy08}
\bibinfo{author}{N.~Halevy}, \bibinfo{title}{Team Negotiation: Social,
  Epistemic, Economic, and Psychological Consequences of Subgroup Conflict},
  \bibinfo{journal}{Personality and Social Psychology Bulletin}
  \bibinfo{volume}{34} (\bibinfo{year}{2008}) \bibinfo{pages}{1687--1702}.

\bibitem[{Thompson et~al.(1996)Thompson, Peterson, and Brodt}]{thompson96}
\bibinfo{author}{L.~Thompson}, \bibinfo{author}{E.~Peterson},
  \bibinfo{author}{S.~Brodt}, \bibinfo{title}{Team negotiation: An examination
  of integrative and distributive bargaining}, \bibinfo{journal}{Journal of
  Personality and Social Psychology} \bibinfo{volume}{70}~(\bibinfo{number}{1})
  (\bibinfo{year}{1996}) \bibinfo{pages}{66--78}.

\bibitem[{Brodt and Thompson(2001)}]{thompson01}
\bibinfo{author}{S.~Brodt}, \bibinfo{author}{L.~Thompson},
  \bibinfo{title}{Negotiating Teams: A levels of analysis},
  \bibinfo{journal}{Group Dynamics} \bibinfo{volume}{5}~(\bibinfo{number}{3})
  (\bibinfo{year}{2001}) \bibinfo{pages}{208--219}.

\bibitem[{Behfar et~al.(2008)Behfar, Friedman, and Brett}]{behfar08}
\bibinfo{author}{K.~Behfar}, \bibinfo{author}{R.~A. Friedman},
  \bibinfo{author}{J.~M. Brett}, \bibinfo{title}{The Team Negotiation
  Challenge: Defining and Managing the Internal Challenges of Negotiating
  Teams}, in: \bibinfo{booktitle}{Proceedings of the 21st Annual Conference for
  the International Association for Conflict Management (IACM-2008)},
  \bibinfo{year}{2008}.

\bibitem[{Ortmann and King(2007)}]{ortmann07}
\bibinfo{author}{G.~Ortmann}, \bibinfo{author}{R.~King},
  \bibinfo{title}{Agricultural cooperatives I: history, theory and problems},
  \bibinfo{journal}{Agrekon} \bibinfo{volume}{46}~(\bibinfo{number}{1})
  (\bibinfo{year}{2007}) \bibinfo{pages}{18--46}.

\bibitem[{Farhangi(2010)}]{smart}
\bibinfo{author}{H.~Farhangi}, \bibinfo{title}{The path of the smart grid},
  \bibinfo{journal}{Power and Energy Magazine, IEEE}
  \bibinfo{volume}{8}~(\bibinfo{number}{1}) (\bibinfo{year}{2010})
  \bibinfo{pages}{18--28}.

\bibitem[{Ramchurn et~al.(2012)Ramchurn, Vytelingum, Rogers, and
  Jennings}]{ramchurn12}
\bibinfo{author}{S.~D. Ramchurn}, \bibinfo{author}{P.~Vytelingum},
  \bibinfo{author}{A.~Rogers}, \bibinfo{author}{N.~R. Jennings},
  \bibinfo{title}{Putting the'smarts' into the smart grid: a grand challenge
  for artificial intelligence}, \bibinfo{journal}{Communications of the ACM}
  \bibinfo{volume}{55}~(\bibinfo{number}{4}) (\bibinfo{year}{2012})
  \bibinfo{pages}{86--97}.

\bibitem[{Brazier et~al.(2002)Brazier, Cornelissen, Gustavsson, Jonker,
  Lindeberg, Polak, and Treur}]{brazier02}
\bibinfo{author}{F.~M. Brazier}, \bibinfo{author}{F.~Cornelissen},
  \bibinfo{author}{R.~Gustavsson}, \bibinfo{author}{C.~M. Jonker},
  \bibinfo{author}{O.~Lindeberg}, \bibinfo{author}{B.~Polak},
  \bibinfo{author}{J.~Treur}, \bibinfo{title}{A multi-agent system performing
  one-to-many negotiation for load balancing of electricity use},
  \bibinfo{journal}{Electronic Commerce Research and Applications}
  \bibinfo{volume}{1}~(\bibinfo{number}{2}) (\bibinfo{year}{2002})
  \bibinfo{pages}{208--224}.

\bibitem[{Lamparter et~al.(2010)Lamparter, Becher, and Fischer}]{lamparter10}
\bibinfo{author}{S.~Lamparter}, \bibinfo{author}{S.~Becher},
  \bibinfo{author}{J.-G. Fischer}, \bibinfo{title}{An agent-based market
  platform for smart grids}, in: \bibinfo{booktitle}{Proceedings of the 9th
  International Conference on Autonomous Agents and Multiagent Systems:
  Industry track}, \bibinfo{organization}{International Foundation for
  Autonomous Agents and Multiagent Systems}, \bibinfo{pages}{1689--1696},
  \bibinfo{year}{2010}.

\bibitem[{Morais et~al.(2012)Morais, Pinto, Vale, and Pra{\c{c}}a}]{morais12}
\bibinfo{author}{H.~Morais}, \bibinfo{author}{T.~Pinto},
  \bibinfo{author}{Z.~Vale}, \bibinfo{author}{I.~Pra{\c{c}}a},
  \bibinfo{title}{Multilevel negotiation in smart grids for VPP management of
  distributed resources}, \bibinfo{journal}{IEEE Intelligent Systems}
  \bibinfo{volume}{27}~(\bibinfo{number}{6}) (\bibinfo{year}{2012})
  \bibinfo{pages}{8--16}.

\bibitem[{Korsgaard et~al.(1995)Korsgaard, Schweiger, and
  Sapienza}]{korsgaard95}
\bibinfo{author}{M.~Korsgaard}, \bibinfo{author}{D.~Schweiger},
  \bibinfo{author}{H.~Sapienza}, \bibinfo{title}{Building commitment,
  attachment, and trust in strategic decision-making teams: The role of
  procedural justice}, \bibinfo{journal}{Academy of Management Journal}
  (\bibinfo{year}{1995}) \bibinfo{pages}{60--84}.

\bibitem[{Sanchez-Anguix et~al.(2011)Sanchez-Anguix, Julian, Botti, and
  Garcia-Fornes}]{sanchez-anguix11}
\bibinfo{author}{V.~Sanchez-Anguix}, \bibinfo{author}{V.~Julian},
  \bibinfo{author}{V.~Botti}, \bibinfo{author}{A.~Garcia-Fornes},
  \bibinfo{title}{{A}nalyzing {I}ntra-{T}eam {S}trategies for {A}gent-{B}ased
  {N}egotiation {T}eams}, in: \bibinfo{booktitle}{10th International Conference
  on Autonomous Agents and Multiagent Systems (AAMAS'11)},
  \bibinfo{pages}{929--936}, \bibinfo{year}{2011}.

\bibitem[{Sanchez-Anguix et~al.(2012{\natexlab{a}})Sanchez-Anguix, Julian,
  Botti, and Garcia-Fornes}]{sanchez-anguix12}
\bibinfo{author}{V.~Sanchez-Anguix}, \bibinfo{author}{V.~Julian},
  \bibinfo{author}{V.~Botti}, \bibinfo{author}{A.~Garcia-Fornes},
  \bibinfo{title}{{R}eaching {U}nanimous {A}greements within {A}gent-{B}ased
  {N}egotiation {T}eams with {L}inear and {M}onotonic {U}tility {F}unctions},
  \bibinfo{journal}{IEEE Transactions on Systems, Man and Cybernetics, Part B}
  \bibinfo{volume}{42}~(\bibinfo{number}{3})
  (\bibinfo{year}{2012}{\natexlab{a}}) \bibinfo{pages}{778--792}.

\bibitem[{Sanchez-Anguix et~al.(2012{\natexlab{b}})Sanchez-Anguix, Dai,
  Semnani-Azad, Sycara, and Botti}]{sanchez-anguix12b}
\bibinfo{author}{V.~Sanchez-Anguix}, \bibinfo{author}{T.~Dai},
  \bibinfo{author}{Z.~Semnani-Azad}, \bibinfo{author}{K.~Sycara},
  \bibinfo{author}{V.~Botti}, \bibinfo{title}{{M}odeling power distance and
  individualism/collectivism in negotiation team dynamics}, in:
  \bibinfo{booktitle}{45 Hawaii International Conference on System Sciences
  (HICSS-45)}, \bibinfo{pages}{628--637}, \bibinfo{year}{2012}{\natexlab{b}}.

\bibitem[{Rubinstein(1982)}]{rubinstein82}
\bibinfo{author}{A.~Rubinstein}, \bibinfo{title}{Perfect equilibrium in a
  bargaining model}, \bibinfo{journal}{Econometrica} \bibinfo{volume}{50}
  (\bibinfo{year}{1982}) \bibinfo{pages}{155--162}.

\bibitem[{Hindriks and Tykhonov(2010)}]{hindriks10}
\bibinfo{author}{K.~V. Hindriks}, \bibinfo{author}{D.~Tykhonov},
  \bibinfo{title}{Towards a quality assessment method for learning preference
  profiles in negotiation}, in: \bibinfo{booktitle}{Agent-Mediated Electronic
  Commerce and Trading Agent Design and Analysis},
  \bibinfo{publisher}{Springer}, \bibinfo{pages}{46--59}, \bibinfo{year}{2010}.

\bibitem[{Marsa-Maestre et~al.(2013)Marsa-Maestre, Klein, Jonker, and
  Aydogan}]{marsa13}
\bibinfo{author}{I.~Marsa-Maestre}, \bibinfo{author}{M.~Klein},
  \bibinfo{author}{C.~M. Jonker}, \bibinfo{author}{R.~Aydogan},
  \bibinfo{title}{From problems to protocols: Towards a negotiation handbook},
  \bibinfo{journal}{Decision Support Systems}
  \doi{\bibinfo{doi}{http://dx.doi.org/10.1016/j.dss.2013.05.019}}.

\bibitem[{Stole(1995)}]{stole95}
\bibinfo{author}{L.~A. Stole}, \bibinfo{title}{Nonlinear pricing and
  oligopoly}, \bibinfo{journal}{Journal of Economics \& Management Strategy}
  \bibinfo{volume}{4}~(\bibinfo{number}{4}) (\bibinfo{year}{1995})
  \bibinfo{pages}{529--562}.

\bibitem[{Bauer et~al.(2001)Bauer, M{\"u}ller, and Odell}]{bauer01}
\bibinfo{author}{B.~Bauer}, \bibinfo{author}{J.~P. M{\"u}ller},
  \bibinfo{author}{J.~Odell}, \bibinfo{title}{Agent UML: A formalism for
  specifying multiagent interaction}, in: \bibinfo{booktitle}{Agent-oriented
  software engineering}, vol. \bibinfo{volume}{1957},
  \bibinfo{organization}{Springer, Berlin}, \bibinfo{pages}{91--103},
  \bibinfo{year}{2001}.

\bibitem[{Gaston and desJardins(2005)}]{gaston05}
\bibinfo{author}{M.~E. Gaston}, \bibinfo{author}{M.~desJardins},
  \bibinfo{title}{Agent-organized networks for dynamic team formation}, in:
  \bibinfo{booktitle}{Proceedings of the fourth international joint conference
  on Autonomous agents and multiagent systems (AAMAS'05)},
  \bibinfo{publisher}{ACM}, \bibinfo{address}{New York, NY, USA}, ISBN
  \bibinfo{isbn}{1-59593-093-0}, \bibinfo{pages}{230--237},
  \bibinfo{year}{2005}.

\bibitem[{Rahwan et~al.(2009)Rahwan, Ramchurn, Jennings, and
  Giovannucci}]{rahwan09}
\bibinfo{author}{T.~Rahwan}, \bibinfo{author}{S.~D. Ramchurn},
  \bibinfo{author}{N.~R. Jennings}, \bibinfo{author}{A.~Giovannucci},
  \bibinfo{title}{An anytime algorithm for optimal coalition structure
  generation}, \bibinfo{journal}{Journal of Artificial Intelligence Research}
  \bibinfo{volume}{34}~(\bibinfo{number}{1}) (\bibinfo{year}{2009})
  \bibinfo{pages}{521--567}, ISSN \bibinfo{issn}{1076-9757}.

\bibitem[{Nurmi(2010)}]{nurmi10}
\bibinfo{author}{H.~Nurmi}, \bibinfo{title}{Voting systems for social choice},
  \bibinfo{journal}{Handbook of Group Decision and Negotiation}
  (\bibinfo{year}{2010}) \bibinfo{pages}{167--182}.

\bibitem[{Russell and Norvig(2003)}]{norvig03}
\bibinfo{author}{S.~Russell}, \bibinfo{author}{P.~Norvig},
  \bibinfo{title}{{Artificial Intelligence: A Modern Approach}},
  \bibinfo{publisher}{Pearson Education}, \bibinfo{year}{2003}.

\bibitem[{Lin et~al.(2012)Lin, Kraus, Baarslag, Tykhonov, Hindriks, and
  Jonker}]{lin12}
\bibinfo{author}{R.~Lin}, \bibinfo{author}{S.~Kraus},
  \bibinfo{author}{T.~Baarslag}, \bibinfo{author}{D.~Tykhonov},
  \bibinfo{author}{K.~Hindriks}, \bibinfo{author}{C.~M. Jonker},
  \bibinfo{title}{Genius: An Integrated Environment for Supporting the Design
  of Generic Automated Negotiators}, \bibinfo{journal}{Computational
  Intelligence}  (\bibinfo{year}{2012}) \bibinfo{pages}{In Press}.

\bibitem[{Baarslag et~al.(2011)Baarslag, Hindriks, and Jonker}]{baarslag11}
\bibinfo{author}{T.~Baarslag}, \bibinfo{author}{K.~V. Hindriks},
  \bibinfo{author}{C.~M. Jonker}, \bibinfo{title}{Towards a Quantitative
  Concession-Based Classification Method of Negotiation Strategies}, in:
  \bibinfo{booktitle}{Agents in Principle, Agents in Practice. Lecture Notes of
  The 14th International Conference on Principles and Practice of Multi-Agent
  Systems}, \bibinfo{pages}{143--158}, \bibinfo{year}{2011}.

\bibitem[{Kawaguchi et~al.(2011)Kawaguchi, Fujita, and Ito}]{kawaguchi11}
\bibinfo{author}{S.~Kawaguchi}, \bibinfo{author}{K.~Fujita},
  \bibinfo{author}{T.~Ito}, \bibinfo{title}{Compromising Strategy Based on
  Estimated Maximum Utility for Automated Negotiation Agents Competition
  (ANAC-10)}, in: \bibinfo{booktitle}{Modern Approaches in Applied
  Intelligence}, vol. \bibinfo{volume}{6704}, \bibinfo{publisher}{Springer
  Berlin / Heidelberg}, \bibinfo{pages}{501--510}, \bibinfo{year}{2011}.

\bibitem[{Baarslag et~al.(2013)Baarslag, Hindriks, and Jonker}]{baarslag13}
\bibinfo{author}{T.~Baarslag}, \bibinfo{author}{K.~V. Hindriks},
  \bibinfo{author}{C.~M. Jonker}, \bibinfo{title}{A Tit for Tat Negotiation
  Strategy for Real-Time Bilateral Negotiations}, vol. \bibinfo{volume}{435} of
  \emph{\bibinfo{series}{Studies in Computational Intelligence}},
  \bibinfo{publisher}{Springer Berlin Heidelberg}, \bibinfo{pages}{229--233},
  \bibinfo{year}{2013}.

\bibitem[{Ehtamo et~al.(2001)Ehtamo, Kettunen, and Hamalainen}]{ehtamo01}
\bibinfo{author}{H.~Ehtamo}, \bibinfo{author}{E.~Kettunen},
  \bibinfo{author}{R.~P. Hamalainen}, \bibinfo{title}{Searching for joint gains
  in multi-party negotiations}, \bibinfo{journal}{European Journal of
  Operational Research} \bibinfo{volume}{130}~(\bibinfo{number}{1})
  (\bibinfo{year}{2001}) \bibinfo{pages}{54--69}.

\bibitem[{Narayanan and Jennings(2006)}]{narayanan06}
\bibinfo{author}{V.~Narayanan}, \bibinfo{author}{N.~R. Jennings},
  \bibinfo{title}{Learning to Negotiate Optimally in Non-stationary
  Environments}, in: \bibinfo{booktitle}{Cooperative Information Agents X, 10th
  International Workshop, CIA 2006, Edinburgh, UK, September 11-13, 2006,
  Proceedings}, vol. \bibinfo{volume}{4149} of \emph{\bibinfo{series}{Lecture
  Notes in Computer Science}}, \bibinfo{publisher}{Springer},
  \bibinfo{pages}{288--300}, \bibinfo{year}{2006}.

\bibitem[{Carbonneau et~al.(2008)Carbonneau, Kersten, and
  Vahidov}]{carbonneau08}
\bibinfo{author}{R.~Carbonneau}, \bibinfo{author}{G.~Kersten},
  \bibinfo{author}{R.~Vahidov}, \bibinfo{title}{Predicting opponent’s moves
  in electronic negotiations using neural networks}, \bibinfo{journal}{Expert
  Systems with Applications} \bibinfo{volume}{34}~(\bibinfo{number}{2})
  (\bibinfo{year}{2008}) \bibinfo{pages}{1266--1273}.

\bibitem[{Carbonneau et~al.(2011)Carbonneau, Kersten, and
  Vahidov}]{carbonneau11}
\bibinfo{author}{R.~A. Carbonneau}, \bibinfo{author}{G.~E. Kersten},
  \bibinfo{author}{R.~M. Vahidov}, \bibinfo{title}{Pairwise issue modeling for
  negotiation counteroffer prediction using neural networks},
  \bibinfo{journal}{Decision Support Systems}
  \bibinfo{volume}{50}~(\bibinfo{number}{2}) (\bibinfo{year}{2011})
  \bibinfo{pages}{449 -- 459}.

\bibitem[{Robu and La~Poutr\'{e}(2008)}]{robu08}
\bibinfo{author}{V.~Robu}, \bibinfo{author}{J.~A. La~Poutr\'{e}},
  \bibinfo{title}{Retrieving the Structure of Utility Graphs Used in Multi-Item
  Negotiation through Collaborative Filtering of Aggregate Buyer Preferences},
  in: \bibinfo{booktitle}{Rational, Robust and Secure Negotiations},
  vol.~\bibinfo{volume}{89} of \emph{\bibinfo{series}{Computational
  Intelligence}}, \bibinfo{publisher}{Springer}, \bibinfo{year}{2008}.

\bibitem[{Williams et~al.(2012)Williams, Robu, Gerding, and
  Jennings}]{williams12}
\bibinfo{author}{C.~R. Williams}, \bibinfo{author}{V.~Robu},
  \bibinfo{author}{E.~H. Gerding}, \bibinfo{author}{N.~R. Jennings},
  \bibinfo{title}{Negotiating Concurrently with Unknown Opponents in Complex,
  Real-Time Domains}, in: \bibinfo{booktitle}{20th European Conference on
  Artificial Intelligence}, vol. \bibinfo{volume}{242},
  \bibinfo{pages}{834--839}, \bibinfo{year}{2012}.

\bibitem[{Mansour and Kowalczyk(2012)}]{mansour12}
\bibinfo{author}{K.~Mansour}, \bibinfo{author}{R.~Kowalczyk}, \bibinfo{title}{A
  Meta-Strategy for Coordinating of One-to-Many Negotiation over Multiple
  Issues}, in: \bibinfo{booktitle}{Foundations of Intelligent Systems}, vol.
  \bibinfo{volume}{122}, \bibinfo{publisher}{Springer Berlin / Heidelberg},
  ISBN \bibinfo{isbn}{978-3-642-25663-9}, \bibinfo{pages}{343--353},
  \bibinfo{year}{2012}.

\end{thebibliography}

\end{document}